\documentclass[11pt]{article}
\usepackage{latexsym}
\usepackage{amssymb}
\usepackage{graphics,graphicx}
\usepackage{tabularx}
\usepackage{amsfonts}
\usepackage{amsmath}
\usepackage{enumerate}
\usepackage{booktabs}
\usepackage{array}

\newcolumntype{L}[1]{>{\raggedright\let\newline\\\arraybackslash\hspace{0pt}}m{#1}}
\newcolumntype{C}[1]{>{\centering\let\newline\\\arraybackslash\hspace{0pt}}m{#1}}
\newcolumntype{R}[1]{>{\raggedleft\let\newline\\\arraybackslash\hspace{0pt}}m{#1}}

\newcommand{\footmsg}[1]{%
  \let\temp\thempfn%
  \def\thempfs{}
  \footnotetext{#1}
  \let\tempfn\temp}

\newcommand{\singlespace} {\baselineskip=12pt \lineskiplimit=0pt \lineskip=0pt}

\newcommand{\beq}{\begin{equation}}
\newcommand{\eeq}{\end{equation}}
\newcommand{\lb}{\label}
\newcommand{\beqar}{\begin{eqnarray}}
\newcommand{\eeqar}{\end{eqnarray}}
\newcommand{\barr}{\begin{array}}
\newcommand{\earr}{\end{array}}

\def\scalp{\mbox{\boldmath$\, \cdot \, $}}

\def\XXint#1#2#3{{\setbox0=\hbox{$#1{#2#3}{\int}$}
     \vcenter{\hbox{$#2#3$}}\kern-.5\wd0}}

\def\inclps#1#2#3{\resizebox{#1}{#2}{\includegraphics{#3}}}

\def\b0{{\bf 0}}

\def\Id{{\bf I}}

\def\bepsilon{\mbox{\boldmath${\epsilon}$}}

\def\bsigma{\mbox{\boldmath${\sigma}$}}

\def\tr{{\rm tr}}

\def\IJF{{\it Int. J. Fracture\ }}

\title{Elastoplastic coupling to model cold ceramic powder compaction}

\author{S. Stupkiewicz$^{(1,2)}$, A. Piccolroaz$^{(1)}$,  and D. Bigoni$^{(1)}$\\[1ex]
\small{(1) -- Department of Civil, Environmental and Mechanical Engineering,}\\
\small{University of Trento, via Mesiano 77, I-38123 Trento, Italy}\\
\small{(2) -- Institute of Fundamental Technological Research (IPPT),}\\
\small{Pawinskiego 5B, 02-106 Warsaw, Poland}\\[1ex]
\small{e-mail: sstupkie@ippt.pan.pl; roaz@ing.unitn.it; bigoni@ing.unitn.it}
}

\date{}

\begin{document}

\maketitle


\begin{abstract}
The simulation of industrial processes involving cold compaction of powders allows for the optimization of the production of both traditional and advanced ceramics. The capabilities of a constitutive model previously proposed by the authors are explored to simulate simple forming processes, both in the small and in the large strain formulation. The model is based on the concept of elastoplastic coupling –providing a relation between density changes and variation of elastic properties– and has been tailored to describe the transition between a granular ceramic powder and a dense green body. Finite element simulations have been compared with experiments on an alumina ready-to-press powder and an aluminum silicate spray-dried granulate. The simulations show that it is possible to take into account friction at the die wall and to predict the state of residual stress, density distribution and elastic properties in the green body at the end of the forming process.

\end{abstract}



\section{Introduction}
\setcounter{equation}{0}

Industrial processes involving the cold forming of ceramic powders can be
numerically simulated to detect the distribution of residual stress, density, and
elastic properties within the green body, with the purpose of improving design
through minimization of defects and regularization of the density distribution. In
this way, the industrial production can be enhanced and consequently rejects and
energy consumption reduced. Simulations, usually based on the finite element
technique, rely on a constitutive model capable of describing the progressive
mechanical densification of a granulate. This is a scientific challenge, as it
involves the description of the transition between two materials with completely
different mechanical behaviour: granular materials are characterized by null
cohesion, \lq drop-shaped' elastic domain, with stress-dependent elastic properties,
while green bodies are cohesive, with a \lq cigar-shaped' elastic domain,
characterized by linear elasticity.

Models developed so far are usually\footnote{A remarkable exception is the study by
Balakrishnan et al. (2011) has been conducted using discrete element simulations.}
variants of known elastoplastic models for soils, based on the \lq cam clay' or \lq
cap' yield surfaces (Aydin et al., 1997a, 1997b; Brandt and Nilsson, 1998, 1999;
Ewsuk et al., 2001a, 2001b; Gu et al., 2006; Henderson et al. 2000; Keller et al.,
1998; Kim et al., 2001; Lee and Kim, 2008; Park and Kim, 2001; Zipse, 1997). In
these models the fact that the elastic stiffness of the green body is a function of
the density, or in other words of the plastic deformation, (Baklouti et al. 1997;
1999; Carneim et al., 2001; Kim et al., 2002; Kounga Njiwa et al., 2006; Zeuch et
al., 2001) is only approximatively kept into account. The key to the rational
incorporation of this effect is the concept of elastoplastic coupling, as initiated
by Hueckel (1975; 1976) and Dougill (1976), and applied by Piccolroaz et al. (2006a)
to the cold densification of ceramic powders, in the form described by Bigoni
(2012).

In addition to coupling between plastic and elastic properties, the model developed
by Piccolroaz et al. (2006a), incorporates micromechanically-based hardening laws
and a previously-defined yield function, called in the following \lq BP' (Bigoni and
Piccolroaz, 2004; Piccolroaz and Bigoni, 2009; Bigoni, 2012). This yield function
possesses the \lq high stretchability' needed to describe the granular/solid
transition typical of ceramics forming, but has the inconvenience that it is defined
infinity outside the elastic domain. This inconvenience, preventing the use of
standard implicit return mapping schemes for integration of rate constitutive
equations, has been recently overcome in different ways (Brannon and Leelavanichkul,
2010; Penasa et al., 2013; Stupkiewicz et al., 2013), so that the model is currently
ready to be used. In particular, computer implementation has been carried out using
the automatic code generation system \emph{AceGen}, and the computations have been
performed using the finite element package \emph{AceFEM} (Korelc, 2002; 2009).

The purpose of the present article is to demonstrate, through a new implementation
allowing us to introduce large strain and friction at the mould/granulate interface,
the capabilities of the above-described model in simulating forming processes of
both advanced (a ready-to-press alumina) and traditional (an aluminum silicate
spray-dried granulate) ceramic powders. To this end, examples of computations are
presented and compared to experiments for uniaxial strain, single and double action
pressing into a mould with friction at the walls and forming with a cross-shaped
punch. The comparison with experiments shows the excellent capability of the model,
particularly in the large strain version, to simulate the mechanical features of
cold forming of ceramic powders.

\section{The concept of elastoplastic coupling}

The force vs.\ displacement relation (with unloadings) for the uniaxial strain of
aluminum silicate spray-dried granulate (6 g powder pressed in a 25 mm diameter
mould), reported in Fig.\ \ref{compattazzo}, shows well-known mechanical features of
cold powder compaction: (i.)\ the large plastic (i.e.\ irreversible) deformation and
(ii.)\ the progressive elastic stiffening of the material, related to the increase
of density.

\begin{figure}[!htcb]
  \begin{center}
     \includegraphics[width= 10 cm]{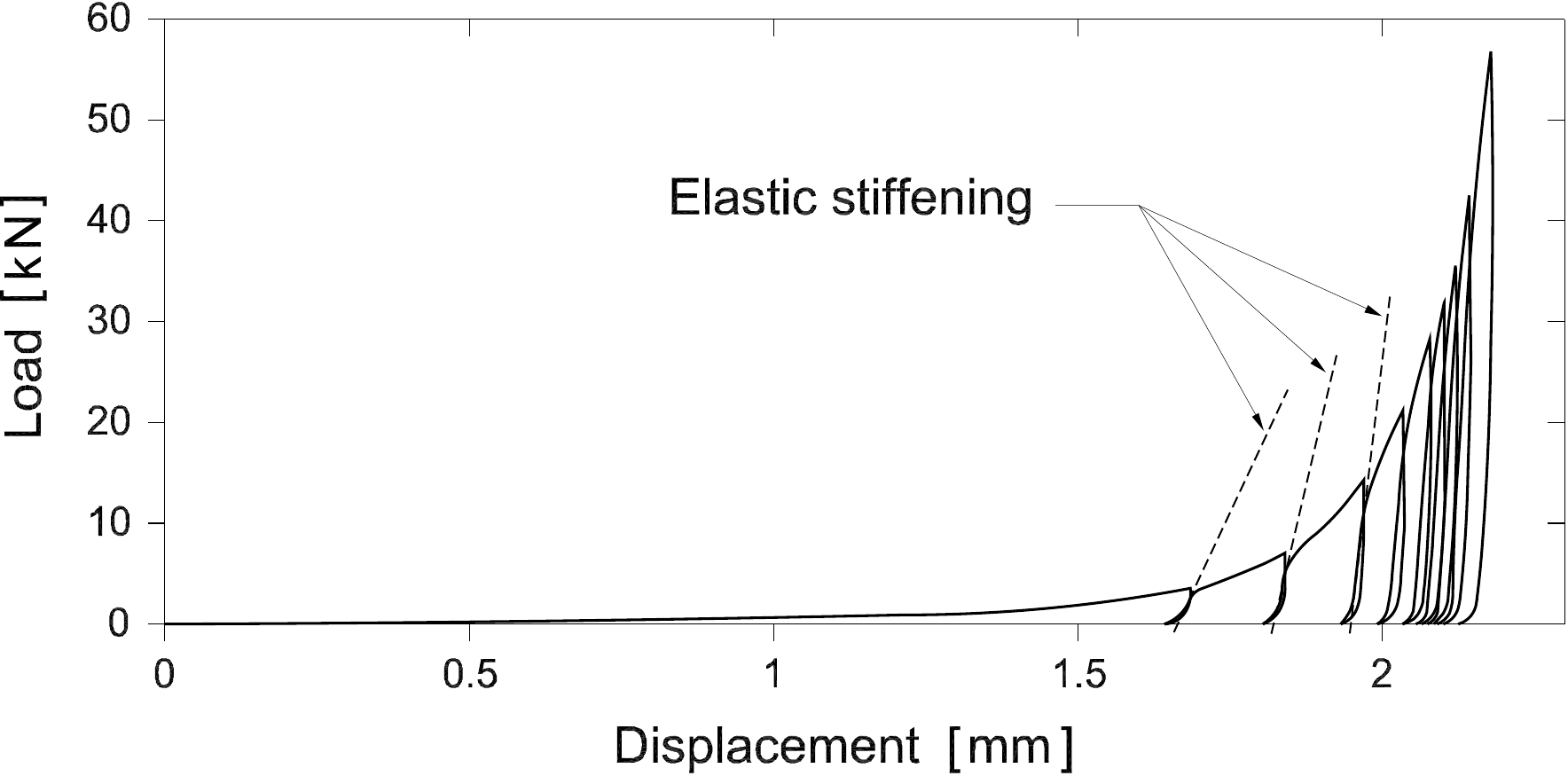}
\caption{\footnotesize Uniaxial deformation of an aluminum silicate spray-dried powder
(force versus displacement) with unloadings, showing the progressive stiffening of the
material.}
\lb{compattazzo}
  \end{center}
\end{figure}

These two features are key aspects in the modelling and can be incorporated in a
constitutive law in a rigorous way through the notion of elastoplastic coupling. In
this section we illustrate this concept with reference to a simple small-strain
model.\footnote{For a comprehensive description of elastoplastic coupling in both
small-strain and finite-strain formulations, the interested reader is referred to
Piccolroaz et al. (2006a, 2006b), Bigoni (2012).}

In particular, the deformation $\epsilon$ is assumed to be the sum of
a plastic, $\epsilon_p$, and an elastic, $\epsilon_e$, component, namely,
\beq
\lb{addittivo}
\epsilon = \epsilon_e+\epsilon_p,
\eeq
where the elastic deformation is induced by the stress $\sigma$, but, due to elastoplastic coupling, is also function of the plastic deformation
\beq
\lb{compliance}
\epsilon_e = \hat{\epsilon}_e(\sigma, \epsilon_p),
\eeq
so that
\beq
\lb{pizza}
\epsilon = \hat{\epsilon}_e(\sigma, \epsilon_p)+\epsilon_p.
\eeq

\noindent
As a response to stress and plastic strain increments,  $\dot{\sigma}$ and $\dot{\epsilon}_p$, the strain is incremented as
\beq
\lb{couplazzo1}
\dot{\epsilon} =
\frac{\partial \hat{\epsilon}_e(\sigma, \epsilon_p)}{\partial \sigma}
\dot{\sigma} +
{\underbrace{\frac{\partial \hat{\epsilon}_e(\sigma, \epsilon_p)}{\partial \epsilon_p}\dot{\epsilon}_{p}}_{e-p~\text{coupling}} }
+ \dot{\epsilon}_p,
\eeq
explicitly showing the elastoplastic coupling term, absent in the usual theories of elastoplasticity, as for instance the Cam-clay.

To elucidate with a simple example the elastoplastic coupled formulation, we refer to the Cooper and Eaton (1962) isostatic compaction model, in
which the plastic volumetric strain $e_v^p$ is related to the hydrostatic compaction pressure $p_c$ as
\beq
e_v^p = -\tilde{a}_1 \exp{\left(-\frac{\Lambda_1}{p_c}\right)} -\tilde{a}_2 \exp{\left(-\frac{\Lambda_2}{p_c}\right)},
\eeq
where, for the aluminum silicate powder reported in Fig.\ \ref{compattazzo}, the constants assume the following values: $\tilde{a}_1$=0.5,
$\tilde{a}_2$=0.05, $\Lambda_1$=0.96, and $\Lambda_2$=40.6.
For the sake of simplicity the following elastic law is adopted
\beq
e_v^e = p_c/B(e_v^p),
\eeq
defined by the elastic bulk modulus $B(e_v^p)$, exponential function of the plastic volumetric deformation
\beq
B(e_v^p) = B^\infty-(B^\infty-B^0)\exp{(-\alpha |e_v^p|)},
\eeq
which can be viewed as a simplification of the law reported by Kim et al. 2002.
The constants $B^\infty$ and $B^0$ assume, for the aluminum silicate powder, the values 4 GPa and 50 KPa, respectively.

Identifying now in eqn (\ref{pizza}) the strain variable with the volumetric strain $e_v$ and the stress variable with $p_c$ we obtain
\beq
\lb{vacca}
e_v = \frac{p_c}{B(e_v^p)} + e_v^p.
\eeq
Eq (\ref{vacca}) describes an isostatic forming process for an aluminum silicate granulate and
has been plotted in Fig.\ \ref{compattazzo_simulato}, which clearly shows the effect of the elastoplastic coupling, namely, an increase of the
elastic stiffness with the increase of the plastic deformation. Note that the model is simple, so that hysteresis of the unloading-reloading behaviour
(visible in Fig.\ \ref{compattazzo}) is not described.

\vspace{3mm}

\begin{figure}[!htcb]
  \begin{center}
     \includegraphics[width= 8 cm]{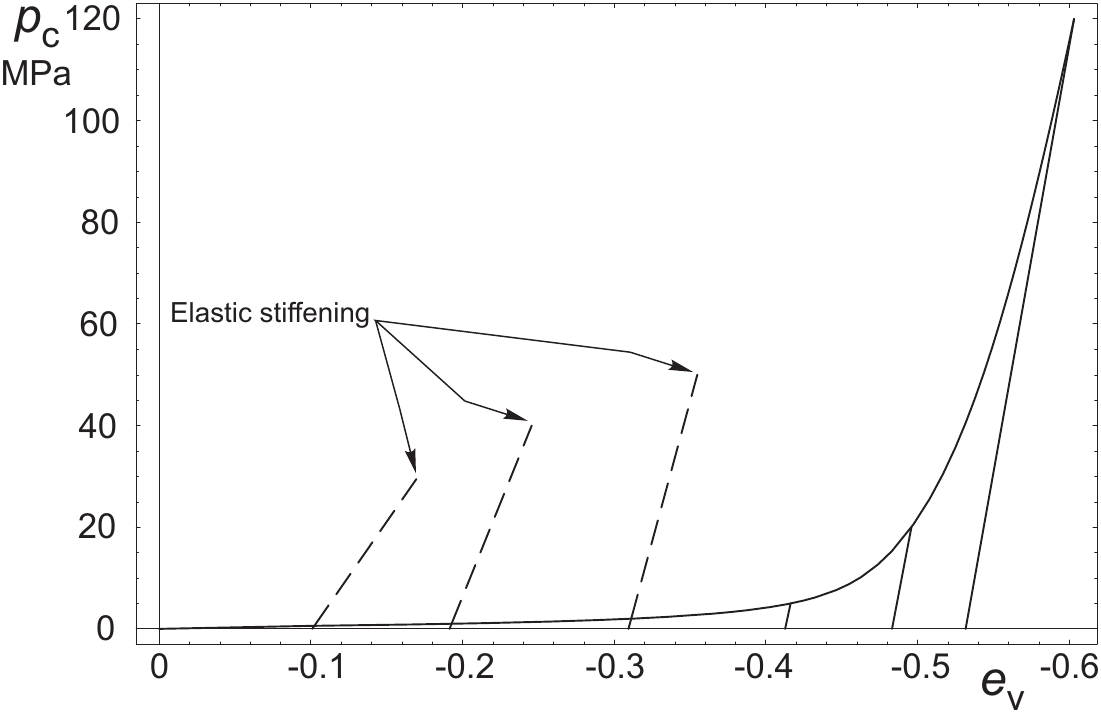}
\caption{\footnotesize Simulated isostatic forming, eq (\ref{vacca}), of an aluminum silicate granulate (compacting pressure versus volumetric deformation)
with unloadings, showing the progressive stiffening of the material due to elastoplastic coupling.
}
\lb{compattazzo_simulato}
  \end{center}
\end{figure}

As an alternative to the compliance approach, eq (\ref{compliance}), a stiffness approach can equivalently be pursued in which the stress
is assumed to be a function of the elastic and plastic strain
\beq
\lb{stiffness}
\sigma = \hat{\sigma}(\epsilon_e, \epsilon_p),
\eeq
so that, as a response to a strain increment $\dot{\epsilon}$, the stress is incremented as
\beq
\dot{\sigma} =
\frac{\partial \hat{\sigma}(\epsilon_e, \epsilon_p)}{\partial \epsilon_e}
\dot{\epsilon}_{e} +
\frac{\partial \hat{\sigma}(\epsilon_e, \epsilon_p)}{\partial \epsilon_p}\dot{\epsilon}_{p},
\eeq
which, using eq (\ref{addittivo}), can be rewritten in a way which explicitly shows the elastoplastic coupling term
\beq
\lb{couplazzo2}
\dot{\sigma} =
\frac{\partial \hat{\sigma}(\epsilon_e, \epsilon_p)}{\partial \epsilon_e}
\dot{\epsilon} -
\frac{\partial \hat{\sigma}(\epsilon_e, \epsilon_p)}{\partial \epsilon_e} \dot{\epsilon}_{p}
{\underbrace{+\frac{\partial \hat{\sigma}(\epsilon_e, \epsilon_p)}{\partial \epsilon_p}
\dot{\epsilon}_{p}}_{e-p~\text{coupling}} }.
\eeq

\noindent
The equivalence of the two formulations (\ref{couplazzo1}) and (\ref{couplazzo2}) can be shown through a substitution of eq (\ref{compliance}) into eq (\ref{stiffness}) to obtain the identity
\beq
\epsilon_e = \hat{\epsilon}_e(\hat{\sigma}(\epsilon_e, \epsilon_p), \epsilon_p),
\eeq
which differentiated gives
\beq
\frac{\partial \hat{\epsilon}_e}{\partial \hat{\sigma}} \frac{\partial \hat{\sigma}}{\partial \hat{\epsilon}_e} = 1, ~~~
-\frac{\partial \hat{\epsilon}_e}{\partial \hat{\sigma}} \frac{\partial \hat{\sigma}}{\partial \epsilon_p}  =
\frac{\partial \hat{\epsilon}_e}{\partial \epsilon_p},
\eeq
allowing the transformation of eq (\ref{couplazzo1}) into eq (\ref{couplazzo2}) and viceversa.

The elastoplastic model developed by Piccolroaz et al. (2006a) for the forming of ceramic powders is based on a stiffness formulation of elastoplastic coupling as that given by eq (\ref{couplazzo2}), where, to enforce hyperelasticity, the stress
\beqar
\lb{elasticlaw}
\lefteqn{\bsigma(\bepsilon_e, e_v^p) =
\left\{ -\frac{2}{3} \mu\, e_v^e + c \right.} \\
& & ~~~~~~~
\left.
+ (p_0 + c)
\left[ \left(d(e_v^p) - \frac{1}{d(e_v^p)}\right) \frac{e_v^e}{\tilde{\kappa}}
- \exp \left( -\frac{e_v^e}{d(e_v^p)^{1/n} \tilde{\kappa}} \right)
\right] \right\} \Id + 2 \mu\, \bepsilon^{e}, \nonumber
\eeqar
(note that $e_v^e = \tr \bepsilon^e$ and $e_v^p = \tr \bepsilon^p$ are respectively the elastic and the plastic volumetric strains) is obtained as the gradient of the following elastic potential
\beqar
\lb{elasticpotential2}
\lefteqn{\phi(\bepsilon^e,e_v^p) =
- \frac{\mu(d)}{3} (e_v^e)^2 + c\, e_v^e} \\
& & ~~~~
+ (p_0 + c) \left[ \left(d - \frac{1}{d}\right) \frac{(e_v^e)^2}{2 \tilde{\kappa}}
+ d^{1/n} \tilde{\kappa}\, \exp \left( - \frac{e_v^e}{d^{1/n} \tilde{\kappa}}  \right) \right]
+ \mu(d)\, \bepsilon^e \scalp \bepsilon^e . \nonumber
\eeqar
In eqs (\ref{elasticlaw}) and (\ref{elasticpotential2}), the parameter $d(e_v^p) \geq 1$ describes the transition
between the logarithmic nonlinear elasticity, typical of granular material (occurring when $d=1$), and the elastic law with stiffness linearly increasing with the forming pressure, typical of the dense green body (occurring when $d \longrightarrow \infty$). In particular,
parameter $d$ is assumed to be a linear function of forming pressure
$p_c$, for values of pressure superior to the breakpoint threshold $p_{cb}$, namely,
\beq
\lb{di}
d = 1 + B <p_c - p_{cb}>,
\eeq
where $B$ is a positive material parameter and
the symbol $<>$ denotes the Macaulay brackets.

In eqs (\ref{elasticlaw}) and (\ref{elasticpotential2})  $\tilde{\kappa}$ is a modification of the logarithmic bulk modulus of the powder, modified through the initial void ratio $e_0$ (at which the mean stress $p$ is equal to $p_0$) as
\beq
\tilde{\kappa} = \kappa / (1 + e_0),
\eeq
$c$ is the cohesion (null in the initial phase of loose granulate), evolving with the following hardening
law
\beq
\lb{ci}
c = c_{\infty} \left[ 1- \exp \left(-\Gamma <p_c - p_{cb}> \right) \right],
\eeq
where $p_{cb}$ is the breakpoint pressure, $c_{\infty}$ and $\Gamma$ are two positive
material parameters, the former
defining the limit value of cohesion reached after large plastic deformation, the latter related
to the
`velocity of growth' of cohesion.
Finally,
the
elastic shear modulus $\mu$ is taken in eqs (\ref{elasticlaw}) and (\ref{elasticpotential2}) to depend on the plastic volumetric strain through the coupling parameter $d$ and the cohesion $c$ as follows
\beq
\lb{mu}
\mu(d) = \mu_0 + c \left( d - \frac{1}{d} \right) \mu_1,
\eeq
where $\mu_0$, and $\mu_1$ are positive material constants.

The final, but essential, ingredient in the modelling of cold powder compaction is the yield surface, defining the stress locus corresponding to the elastic range, which in the form introduced by Bigoni and Piccolroaz (2004) is given by
\beq
\lb{yieldfunction}
f(p, p_c, c) + \frac{q}{g(\theta)} = 0,
\eeq
where $q=\sqrt{3\bsigma\scalp\bsigma/2 -\tr^2\bsigma/2}$ is the second deviatoric invariant, $f(p, p_c, c)$ is the function defining the dependence on the mean  pressure $p = -\tr\bsigma/3$, namely,
\beq
\lb{effedip}
f(p, p_c, c) =
-M p_c \sqrt{\left[\frac{p + c}{p_c + c} - \left(\frac{p + c}{p_c + c} \right)^m\right]\left[2 (1 - \alpha) \frac{p + c}{p_c + c} + \alpha\right]},
\eeq
and $g(\theta)$ describes dependence on the Lode's invariant $\theta$, assumed in the form
\beq
\lb{gi}
g(\theta) = \frac{1}{\cos{\left[ \beta \frac{\pi}{6} -
\frac{1}{3} \cos^{-1} \left(\gamma \cos{3 \theta}\right)\right]}}.
\eeq
The shape of the yield surface depends on the six parameters $\alpha, M, p_c, c, \beta, \gamma$ and is extremely flexible and thus capable of describing the yield locus of many different materials.
The yield surfaces for the ready-to-press alumina and for the aluminum silicate granulate employed in the present article are shown in Fig. \ref{figBP}.

\vspace{3mm}

\begin{figure}[!htcb]
  \begin{center}
     \includegraphics[width= 8 cm]{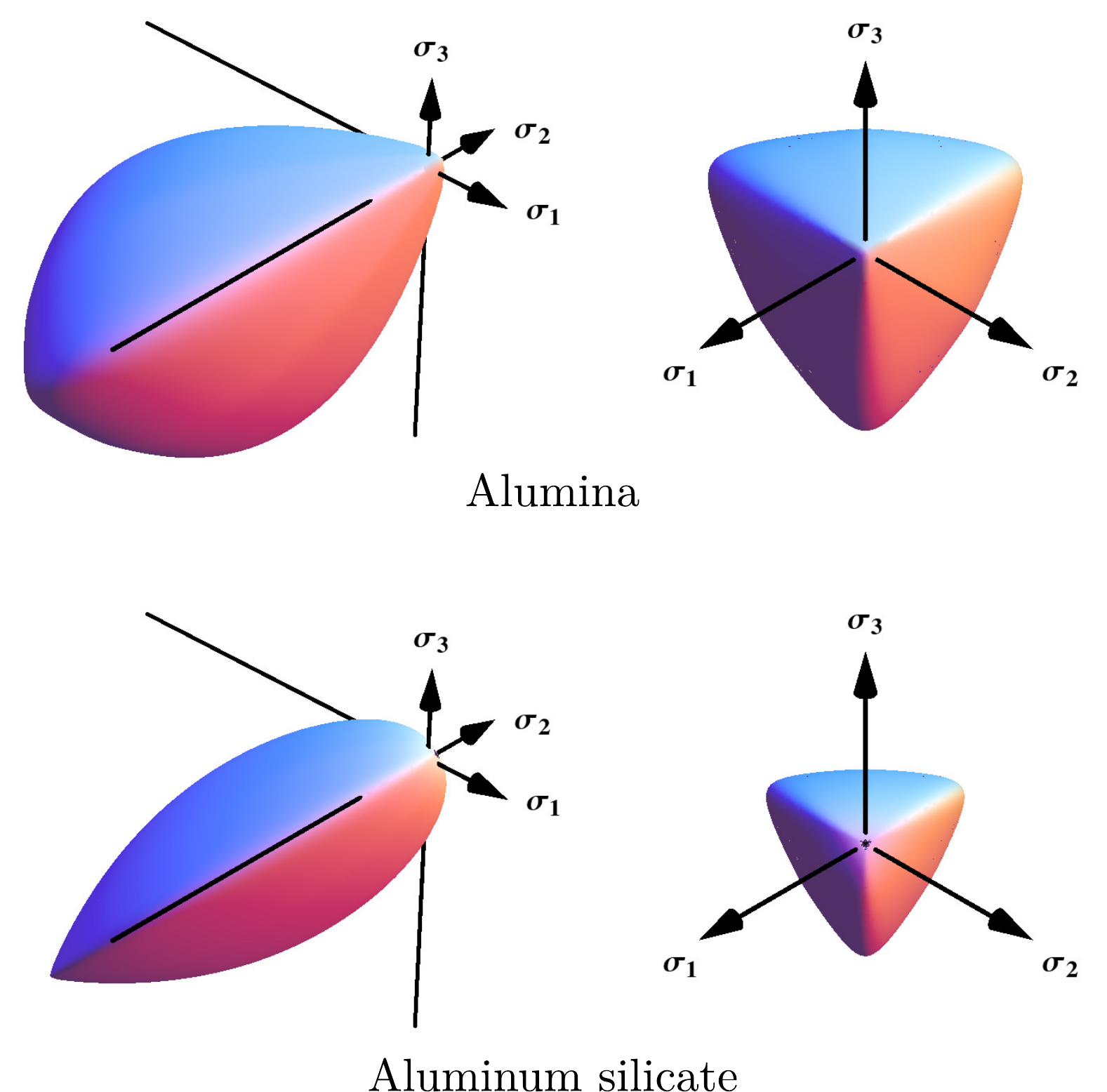}
\caption{\footnotesize The BP yield surfaces with the parameters suited to model the ready-to-press alumina powder and the aluminum silicate granulate employed in the present article.
}
\lb{figBP}
  \end{center}
\end{figure}

The described elastoplastic coupled model has been also produced in a large strain
version by Piccolroaz et al. (2006b). Both the small and the large strain versions
of the model have been implemented in a finite element program and will be employed
in the next section to assess their predictive capabilities.

\section{Numerical simulations versus experimental results}

Simulations of cold compaction processes are presented in this section with
reference to:
\begin{itemize}
\item the ready-to-press 392 Martoxid KMS-96 alumina powder characterized by the
    constitutive parameters identified by Piccolroaz et al. (2006a, their Table
    1), with the slight modifications reported below;
\item the aluminum silicate spray-dried powder (used in the production of tiles)
    experimentally characterized by Bosi et al. (2013).
\end{itemize}
The simulations, performed with the small and the large strain versions of the
elastoplastic coupled model presented in the previous section, refer to uniaxial
deformation, single and double action compaction in a mould with friction at the
powder/wall contact, and forming into a mould with a cross-shaped punch, designed by
us to induce a strong variation in the mechanical properties after forming. In all
cases the simulations have been compared to experimental results, some already
available (Piccolroaz et al. 2006a; Bosi et al., 2013) and other performed for the
present study.

In order to enable direct comparison of the small and the large strain
models, the constitutive parameters adopted for the alumina powder have been
slightly modified with respect to the values reported by Piccolroaz et al.\ (2006a)
so that the uniaxial response of the two models is as close as possible. This is
illustrated in Section \ref{sec:uniaxial}. Specifically, parameter $B$ has been set
to $B=0.6$ MPa$^{-1}$ in the small strain model and to $B=0.18$ MPa$^{-1}$ in the
finite strain model, while parameters $\tilde{a}_1$ and $\tilde{a}_2$ have been set
to $\tilde{a}_1=0.383$ and $\tilde{a}_2=0.124$ for both models (compared to the
original values of 0.37 and 0.12, respectively). Further, parameter $\epsilon$
defining the flow rule non-associativity has been set to $\epsilon=0.5$.
In the case of the aluminum silicate powder, the parameter $M$ of the yield surface has been increased to 0.5 with respect to the value reported by Bosi et al. (2013) to avoid 
near-boundary numerical instabilities. 
Material parameters of the models employed for the finite strain simulations are summarized
in Table \ref{tab01}, where $\rho_T$ is the teoretical density of the powder. 


\begin{table}[!htcb]
\newcommand{\ra}[1]{\renewcommand{\arraystretch}{#1}}
\small
\centering
\ra{1.2}
\begin{tabular}{@{}L{35mm} C{19mm} C{21mm} C{18mm} C{15mm} C{15mm}@{}}
\toprule
Yield surface           & $M$              & $m$               & $\alpha$         & $\beta$        & $\gamma$       \\
\cmidrule(l){2-6}
Alumina                 &  1.1             &  2                &  0.1             &  0.19          &  0.9           \\
Aluminum silicate       &  0.5         &  4.38             &  1.95            &  0.1           &  0.9           \\
\midrule
Elastic log. bulk mod.  &  $\kappa$        &                   &   Flow rule      &   $\epsilon$   &                \\
\cmidrule(l){2-2}  \cmidrule(l){5-5}
Alumina                 &  0.04            &                   &                  &  0.5           &                \\
Aluminum silicate       &  0.08            &                   &                  &  0.5           &                \\
\midrule
1st hardening law       & $\tilde{a}_1$    & $\tilde{a}_2$     & $\Lambda_1$      & $\Lambda_2$    &                \\
\cmidrule(l){2-5}
Alumina                 &  0.383           &  0.124            &  1.8 MPa         &  40 MPa        &                \\
Aluminum silicate       &  0.497           &  0.057            &  1.14 MPa        &  40.9 MPa      &                \\
\midrule
2nd hardening law       & $c_\infty$       & $\Gamma$          & $p_{cb}$         &                &                \\
\cmidrule(l){2-4}
Alumina                 &  2.3 MPa         &  0.026 MPa$^{-1}$ &  3.2 MPa         &                &                \\
Aluminum silicate       &  0.778 MPa       &  0.046 MPa$^{-1}$ &  1.3 MPa         &                &                \\
\midrule
E-P coupling            & $B$              & $n$               & $\mu_0$          & $\mu_1$        &                \\
\cmidrule(l){2-5}
Alumina                 &  0.18 MPa$^{-1}$ &  6                &   1 MPa          &    64          &                \\
Aluminum silicate       &  2.16 MPa$^{-1}$ &  6                &   1 MPa          &    16          &                \\
\midrule
Initial state           & $e_0$              & $p_0$           & $\rho_T$         &                &                \\
\cmidrule(l){2-4}
Alumina                 &   2.129        & 0.063 MPa         &   3.98 g/cm$^3$    &                &                \\
Aluminum silicate       &   1.741         &  0.09 MPa       &  2.60   g/cm$^3$    &               &                \\
\bottomrule
\end{tabular}
\caption{Material parameters used for finite strain simulations.}
\label{tab01}
\end{table}

The present finite element implementation of the model, with full account
for the elastoplastic coupling, employs the standard return mapping algorithm
combined with the implicit backward-Euler time integration scheme. The difficulties
associated with the application of the return mapping algorithm to the BP yield
surface have been overcome in different ways, one reported by Penasa et al. (2013), and the other used
here is based on the concept of implicit yield surface that
is described by Stupkiewicz et al. (2013). The computer implementation has
been carried out using the automatic code generation system \emph{AceGen} (Korelc,
2002; 2009). In particular, the automatic differentiation technique implemented in
\emph{AceGen} has been applied to derive the consistent tangent matrix. Exact
linearization of the incremental constitutive relationships is crucial for the
convergence of the Newton method used to solve the nonlinear finite element
equations.

The example studied in Section \ref{sec:friction} involves frictional
contact interaction of the ceramic powder sample with the mould and the punch.
Efficient and robust treatment of the corresponding unilateral contact and friction
constraints relies on application of the augmented Lagrangian method of Alart and
Curnier (1991), see also Lengiewicz et al.\ (2011) for the details of the computer
implementation.

\subsection{Uniaxial strain of ceramic powder}
\label{sec:uniaxial}

\begin{figure}[!htcb]
  \centerline{\inclps{0.95\textwidth}{!}{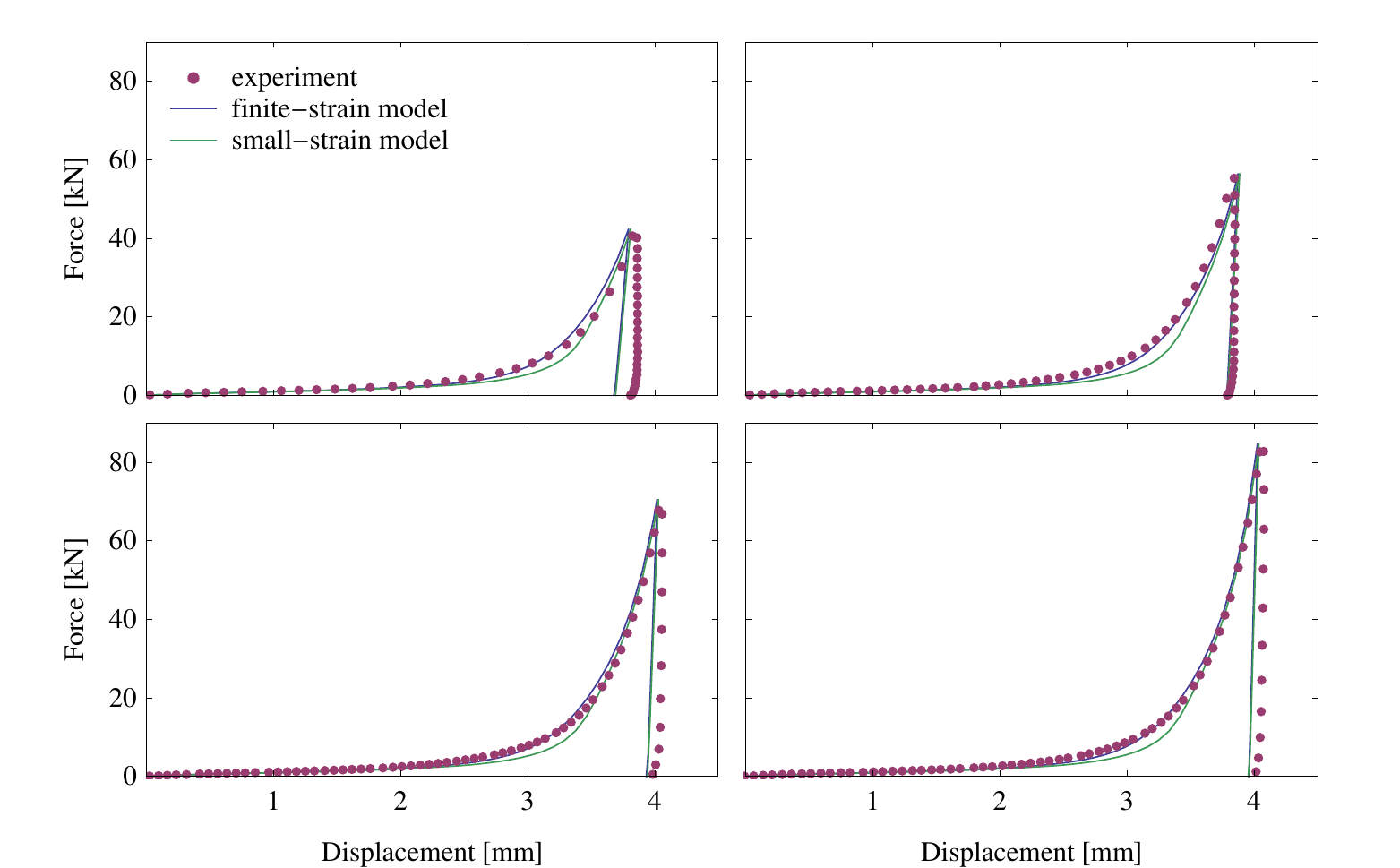}}
  \caption{Uniaxial strain of alumina: small-strain and finite-strain models
           compared to the experiment of Piccolroaz et al. (2006a). Upper part:
           compaction at 60MPa (left) and at 80MPa (right). Lower part:
           compaction at 100MPa (left) and at 120MPa (right).
           \label{fig:a:1}}
\end{figure}

Simulations of uniaxial strain of alumina powder are reported and compared to
experimental results in Fig. \ref{fig:a:1}. The results are reported in terms of
force versus displacement relations and referred to four 30 mm diameter cylindrical
specimens of the nominal initial height of 8.7 mm compacted to 60, 80, 100 and 120
MPa and subsequently unloaded. The actual initial height of the specimens was 8.73,
8.64, 8.78 and 8.66 mm, respectively.

Simulations show an excellent agreement with experimental results, particularly for
the large strain analyses. As discussed above, the constitutive parameters
of the small and the large strain models have been adopted such that the uniaxial
response of the two models is as close as possible. The two responses are indeed
very close except in an intermediate range of small compaction pressures where a small
difference is observed, see Fig. \ref{fig:a:1}.

\subsection{Single and double action forming into a mould with wall friction: density distribution}
\label{sec:friction}

We simulate single and double action cold compression of ceramic powder into a (38.2
mm diameter) cylindrical rigid mould in the presence of Coulomb friction between the
powder and the mould wall, with the friction coefficient equal to 0.4. Samples of
alumina and aluminum silicate have been analyzed with an initial height of 113.9 mm
and 128.5 mm, respectively. Numerical simulation (to mimick the experimental
procedure) has been performed providing first a frictionless preforming uniaxial
strain corresponding to a compression of 3 MPa and later providing the final forming
pressure of 40 MPa in the presence of friction. The problem is treated as
axisymmetric.

\begin{figure}[!htcb]
  \centerline{\inclps{0.50\textwidth}{!}{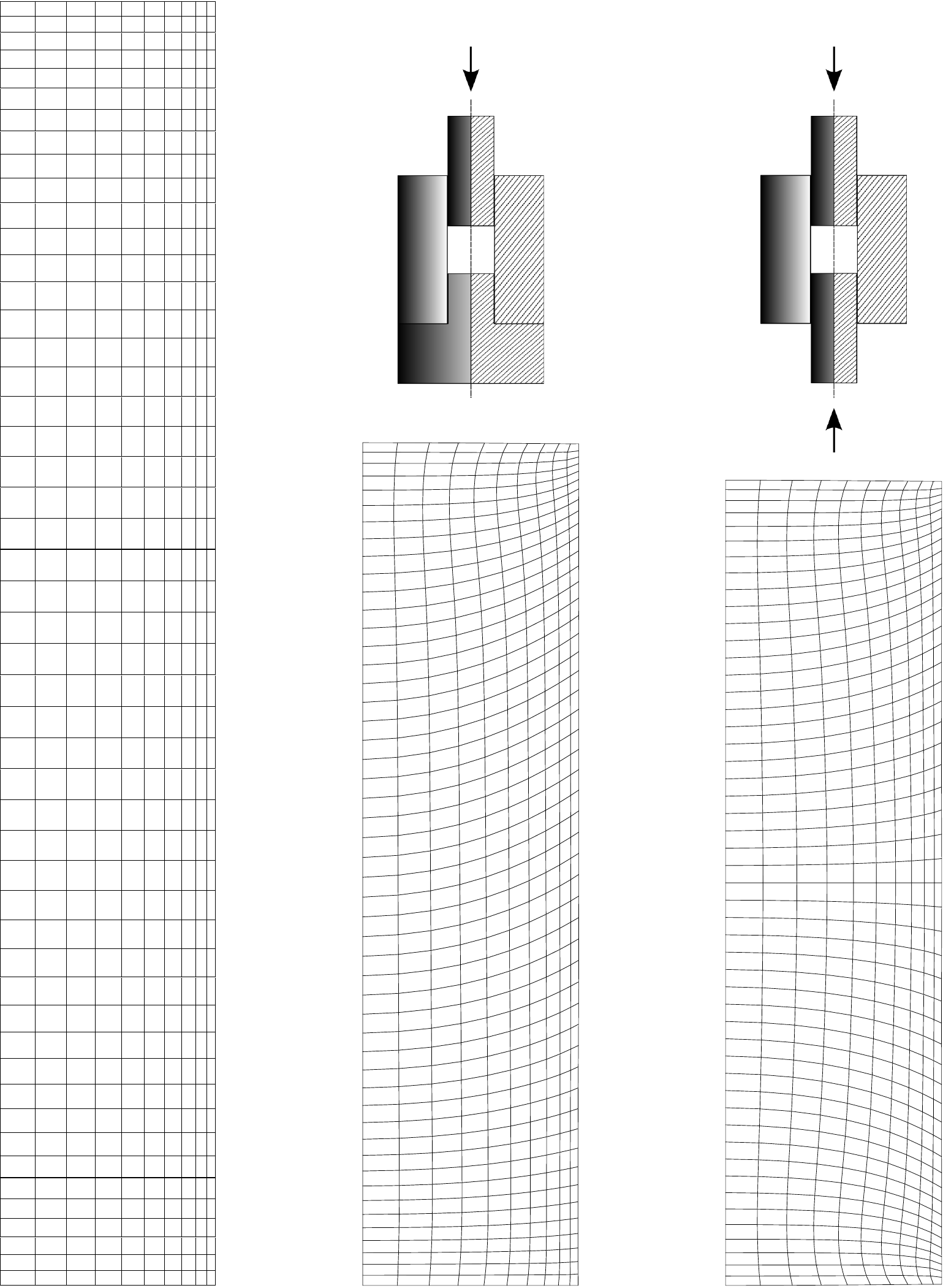}}
  \caption{Finite element mesh for single and double action forming of alumina powder with friction (coefficient has been assumed equal to 0.4) at the mould wall. From left to right: undeformed, deformed for single action compression, and for
           double action compression.
           \label{fig:3}}
\end{figure}

Undeformed and deformed (after loading, subsequent unloading, and
ejection from the mould) meshes are reported in Fig. \ref{fig:3} for single action
and double action compression of alumina powder.
In addition to the nonuniform mesh distortion induced by friction at the wall, we
can note from Fig. \ref{fig:3} the better compaction reached with double action,
leading to a final height smaller than that obtained with the single action
compression. 

\begin{figure}[!p]
  \centerline{
    \begin{tabular}{ccccc}
      \inclps{!}{0.5\textwidth}{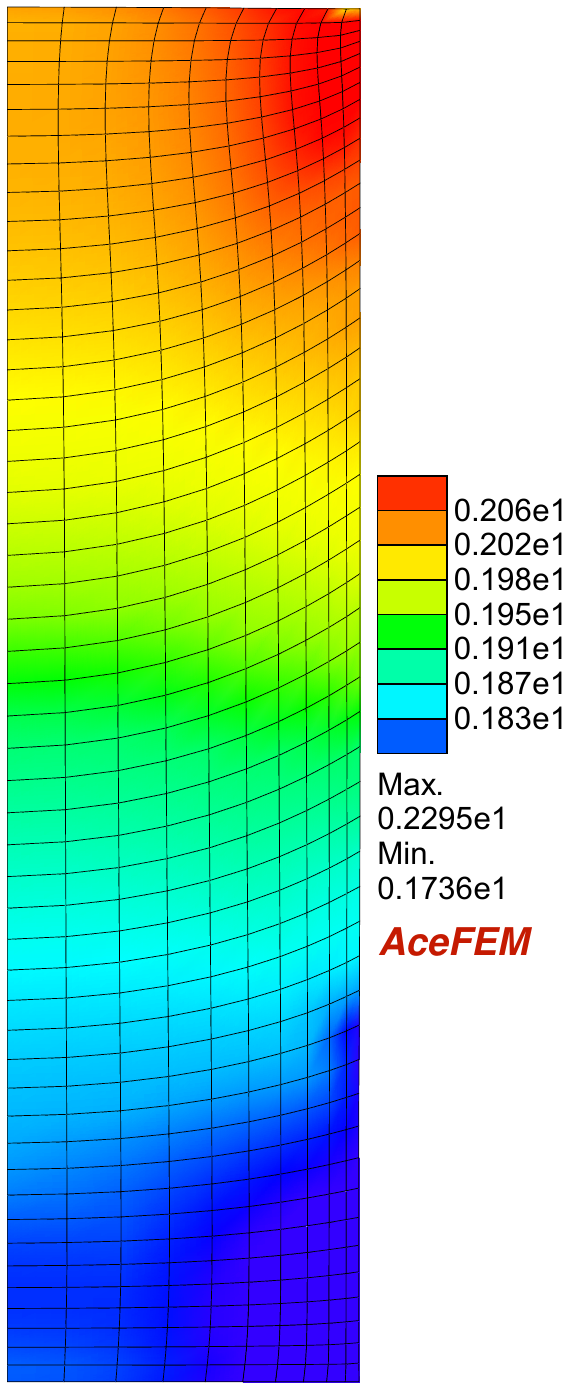} & &
      \inclps{!}{0.5\textwidth}{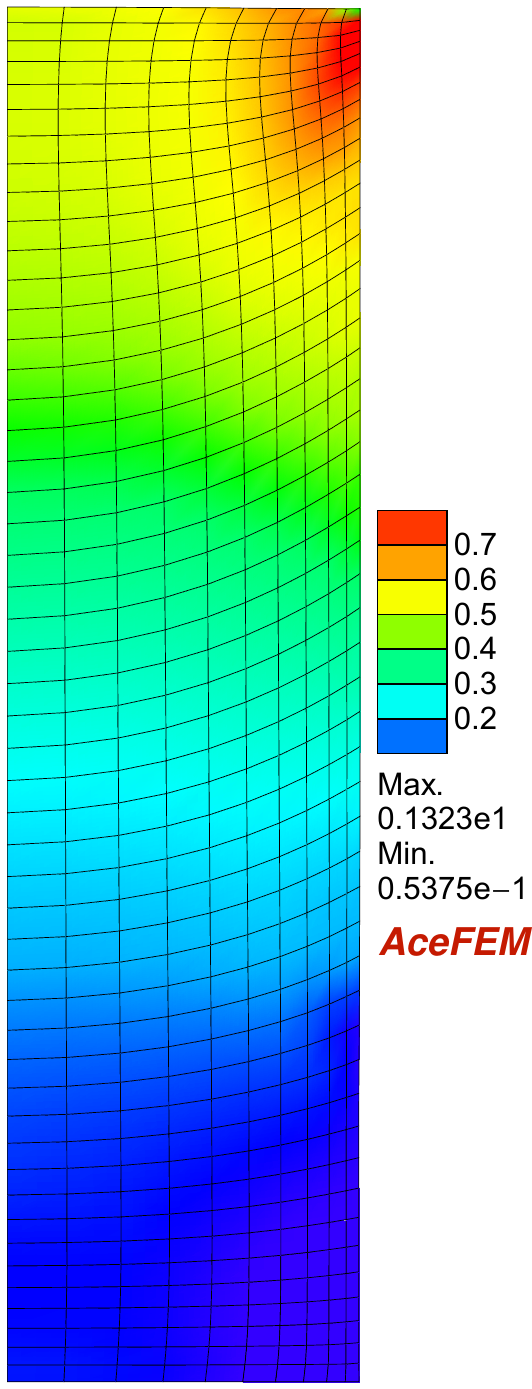} & &
      \inclps{!}{0.5\textwidth}{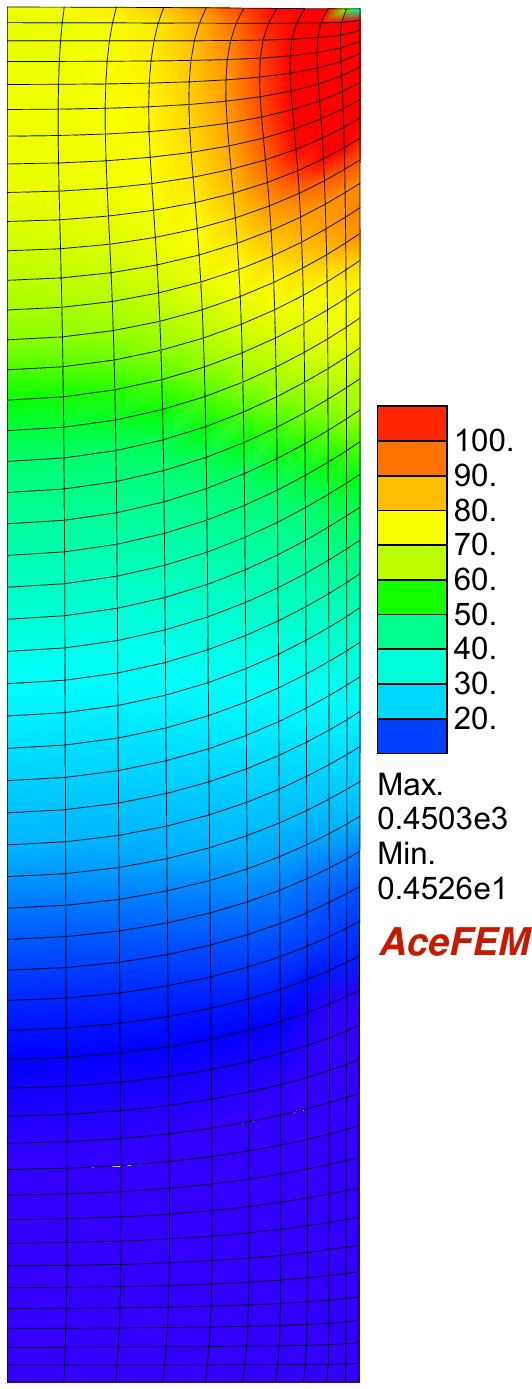} \\[2ex]
      \inclps{!}{0.47\textwidth}{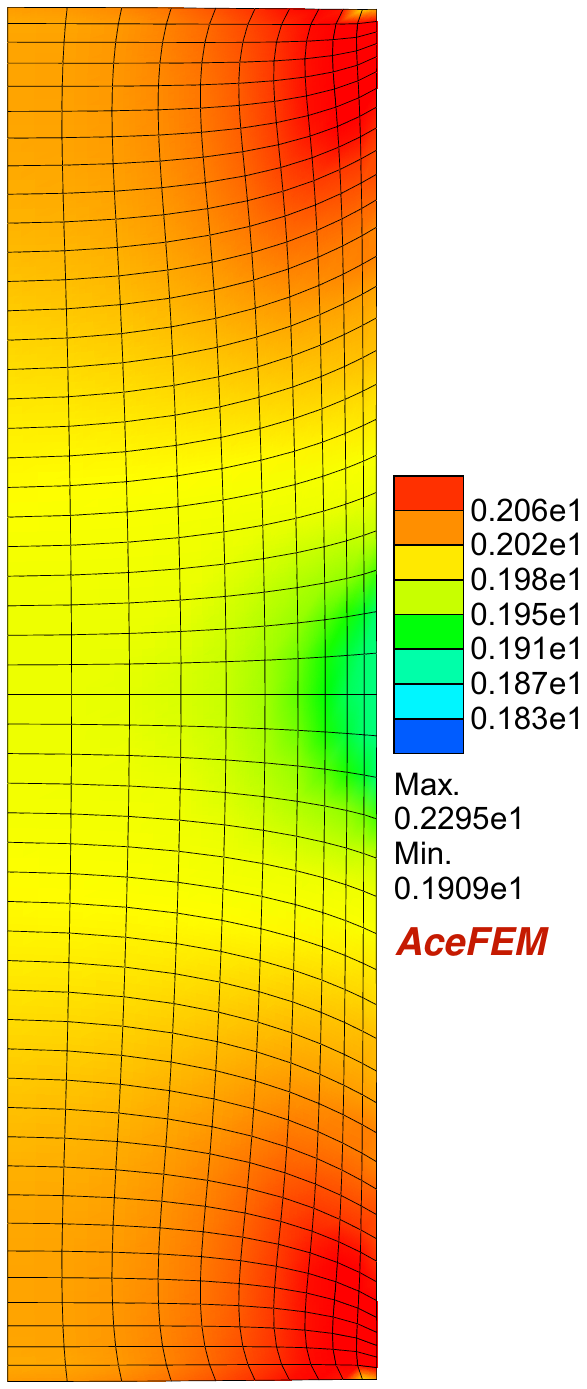} & &
      \inclps{!}{0.47\textwidth}{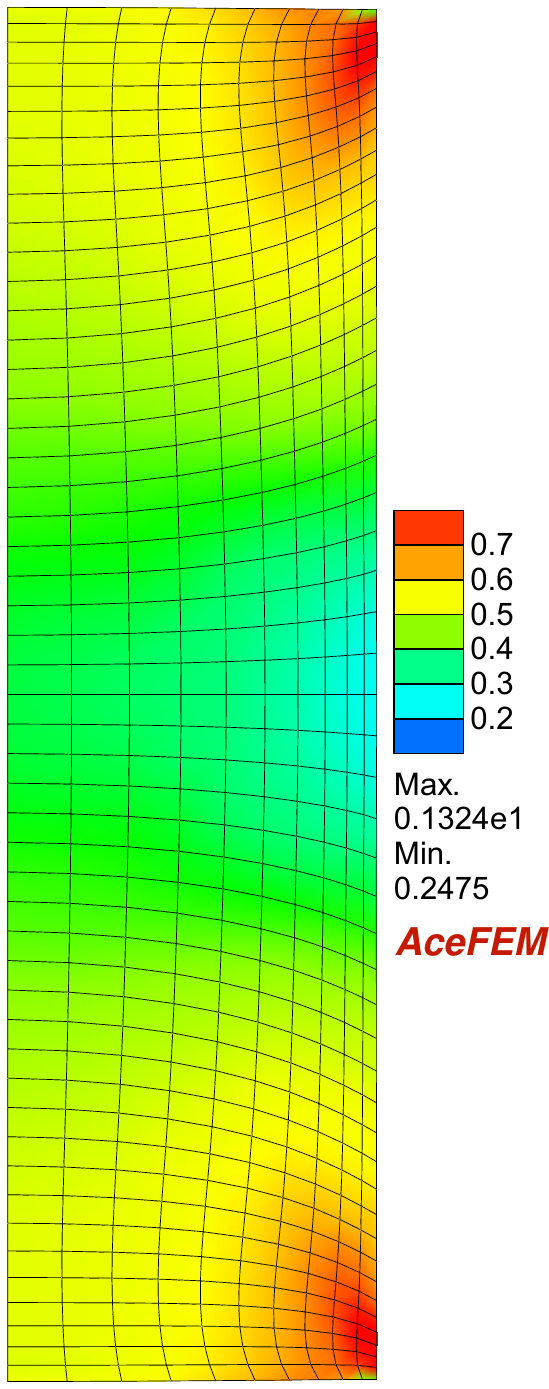} & &
      \inclps{!}{0.47\textwidth}{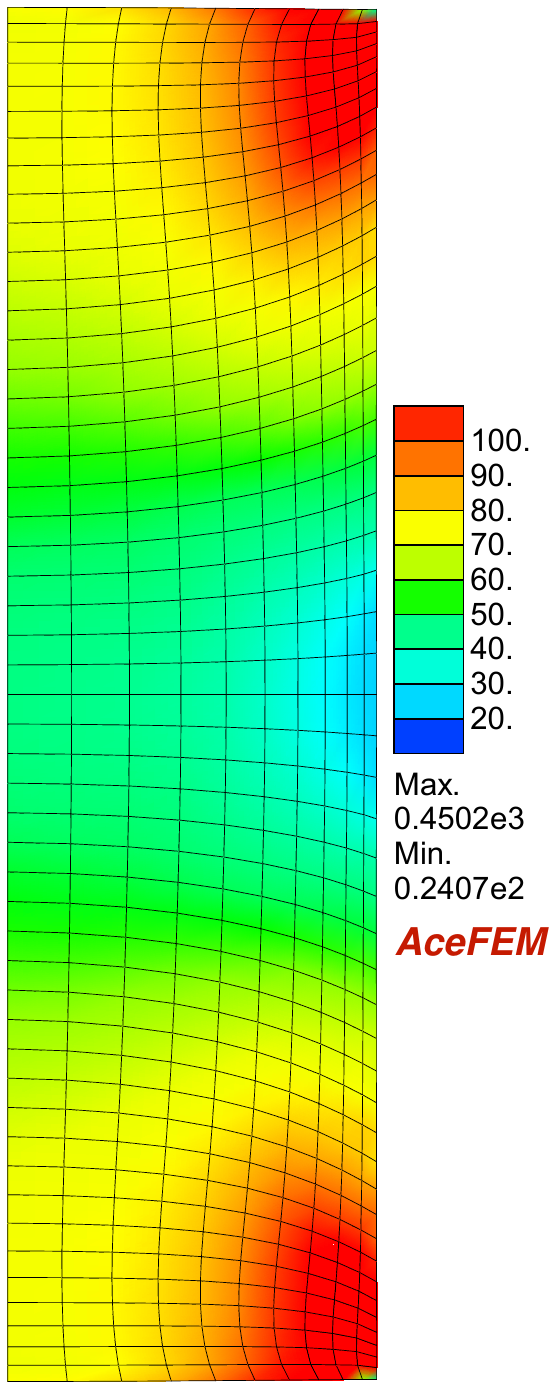}
    \end{tabular}
    }
  \caption{Distribution of density $\varrho$ (in g/cm$^3$, left), cohesion $c$ (in MPa, centre), tangent
           elastic bulk modulus $K_t$ (in MPa, right) for single action compression (top row) and
           double action compression (bottom row) of alumina powder (wall/powder
           and punch/powder friction coefficient was assumed equal to 0.4).
           \label{fig:4}}
\end{figure}

The distributions of density, cohesion and tangent elastic bulk modulus
are shown in Fig. \ref{fig:4} for single (upper part) and double (lower part) action
compression.
This figure clearly shows that density, cohesion and elastic properties obtained at
the end of the forming process are much more uniform with the double action device
than with the single action.

\begin{figure}[!htcb]
  \centerline{
    \begin{tabular}{cc}
      \inclps{0.45\textwidth}{!}{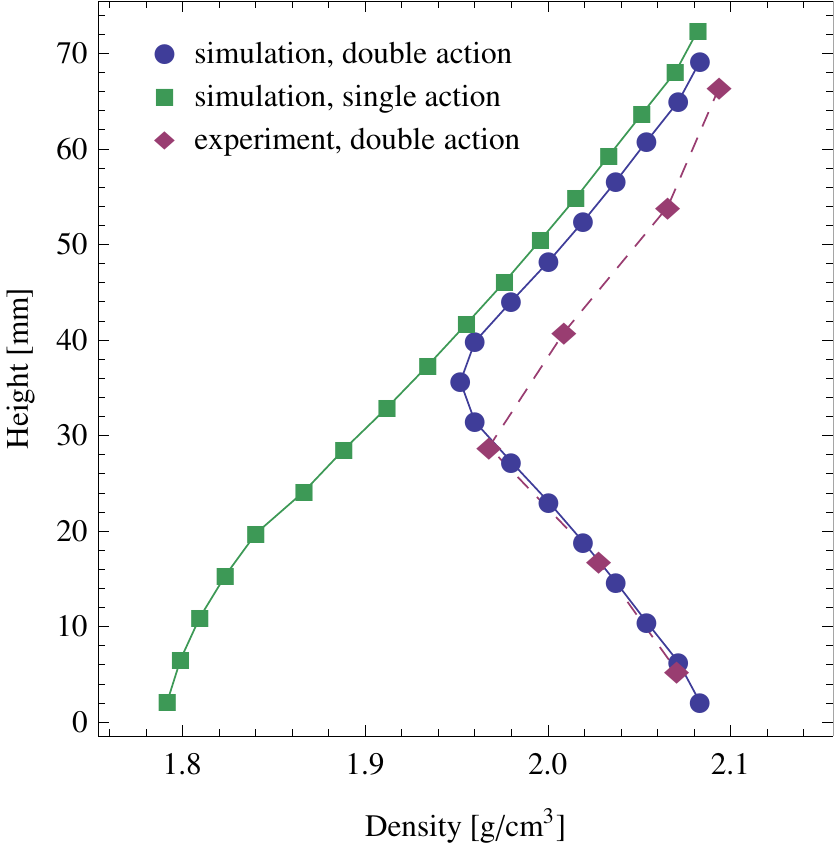} &
      \inclps{0.45\textwidth}{!}{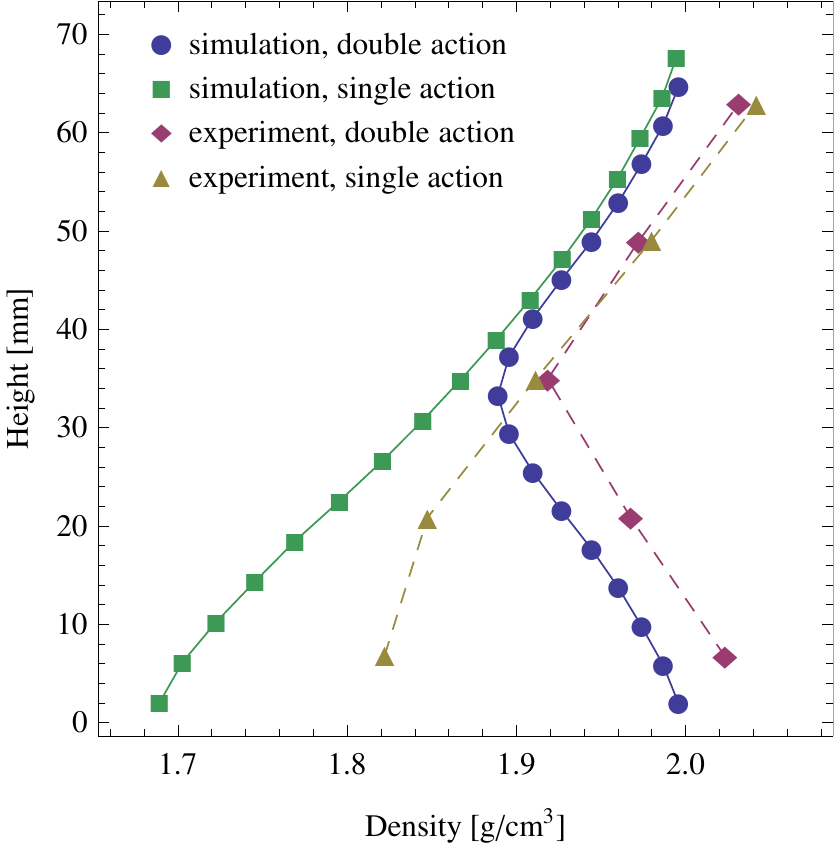}
    \end{tabular}
    }
  \caption{Density distribution for single and double action compression
           in a mould with friction at the wall: alumina (left) and aluminum
           silicate (right) powder. Experimental results are also included.
           \label{fig:1}}
\end{figure}

Finally, the density distribution (averaged through the cross section of the sample) along the height of the
sample, simulated for the alumina and the aluminum silicate powder, are reported in
Fig. \ref{fig:1} and compared with experimental results. Experiments have been
performed by subsequently charging a mould with powder and compressing at 3 MPa, so
that to obtain a \lq layered' sample, to be finally loaded at 40 MPa. Due to the
small height of the \lq layering', the influence of friction was
negligible during the compression at 3 MPa, but friction was important
during compaction at 40 MPa.

Note that experiments for the single action compression were not performed on the
alumina sample (so that these are not reported in Fig. \ref{fig:1}). The simulated
density distributions are in good qualitative and quantitative agreement with
experimental results, with a discrepancy partly due to the fact that
the double action compression was simulated assuming perfectly symmetric action of
the two punches while this symmetry was not exactly preserved in the experiment.

\newpage

\subsection{Forming in a mould with a cross-shaped punch}

\begin{figure}[!htcb]
  \centerline{
       \inclps{0.5\textwidth}{!}{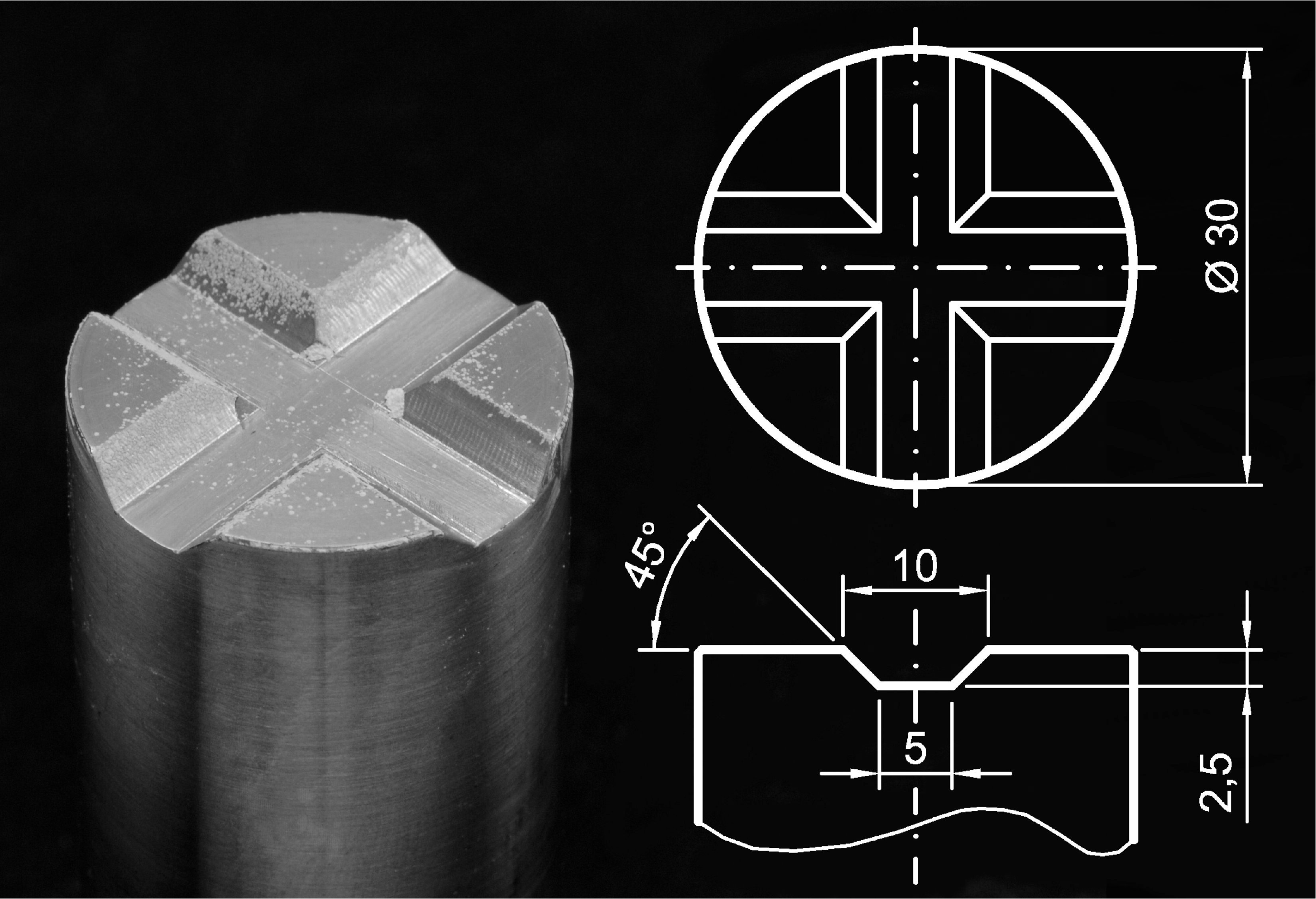}
    }
  \caption{The cross-shaped punch (quotes are in mm) used to form the pieces shown in Fig. \ref{intatta}.
           \label{punzone}}
\end{figure}

To induce large inhomogeneities during cold forming of a sample, we have designed
and manufactured a (30 mm diameter) cross-shaped punch with 2.5 mm deep grooves,
shown in Fig. \ref{punzone}, where the quotes are in mm.

\begin{figure}[!htcb]
  \centerline{
       \inclps{0.38\textwidth}{!}{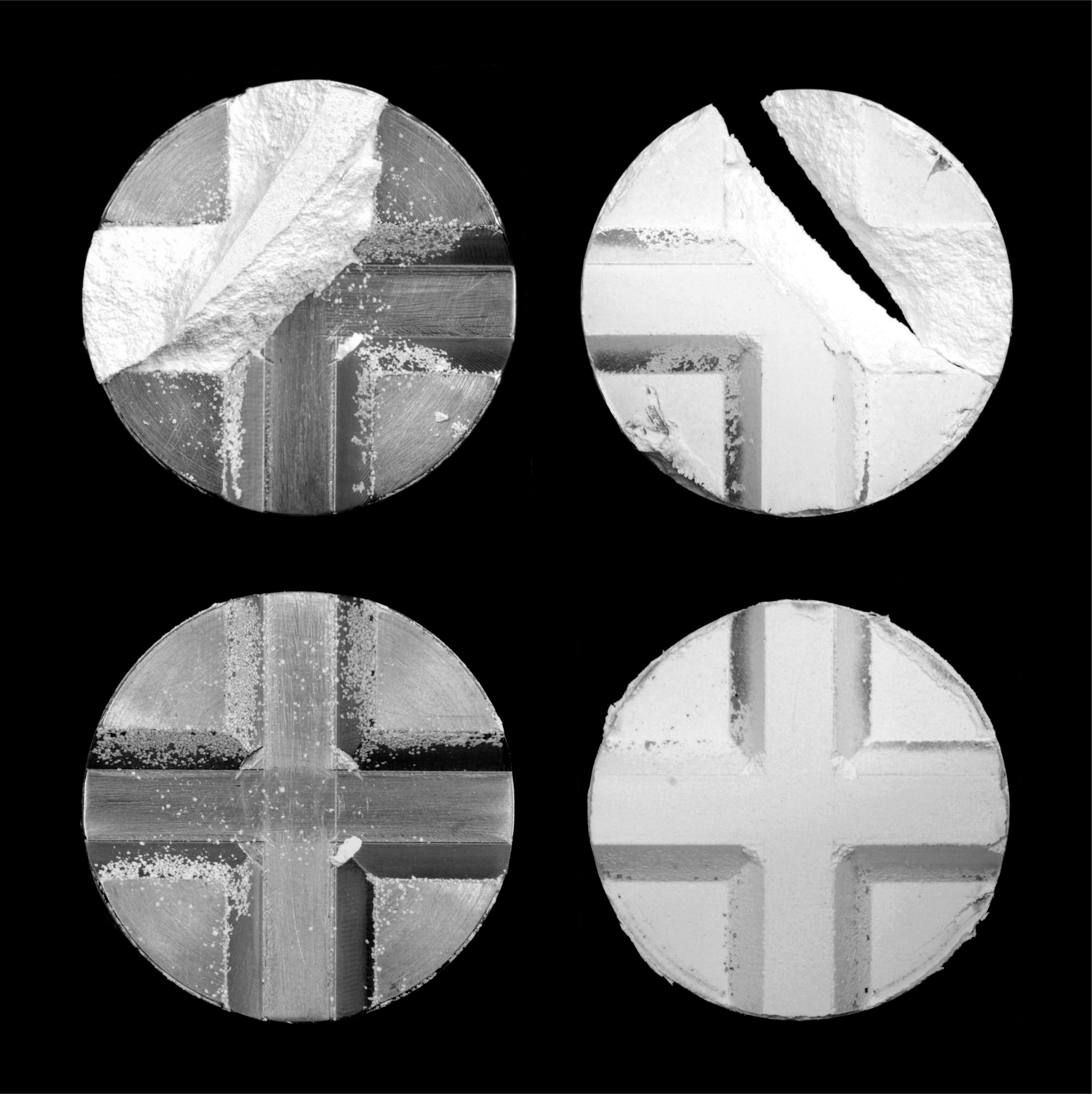}
    }
  \caption{Cross-shaped (30 mm diameter) punch (left) and the formed pieces (right), broken (upper part) and intact (lower part) after ejection.
           \label{intatta}}
\end{figure}

The punch has been used to form pieces from 8 g of alumina powder at a vertical load
of 70 kN, see Fig. \ref{intatta}.
As a consequence of the highly inhomogeneous density distribution obtained with the
cross-shaped punch, the sample sometimes breaks after mold ejection, see upper part of Fig.
\ref{intatta}.

By exploiting the symmetry of the punch, only one quarter of the specimen
has been modelled in order to reduce the computation time. Simplified boundary conditions
are applied as described below. The loading is applied by prescribing the vertical
displacement at the top surface, while the horizontal displacements at the top
surface are fully constrained. At the bottom surface, all displacements are
constrained. This corresponds to high friction and sticking contact at the top and
the bottom surfaces. The horizontal displacements are constrained on the lateral
surface which approximately corresponds to frictionless contact at the mould
surface. Considering that the aspect ratio of the specimen is low, friction at the
mould surface is expected to have a small influence on the forming force and on the
deformation pattern.

Unloading and spring-back are modelled in two steps. In the first step, the reaction
forces at the top and bottom surfaces are gradually decreased to zero to simulate
unloading of the punch. In the second step, the reaction forces at the lateral surface are gradually
decreased to zero to simulate ejection from the mould. Note that both steps are
accompanied by plastic deformations induced by specimen inhomogeneity and the
associated residual stresses.

\begin{figure}[!htcb]
  \centerline{
    \begin{tabular}{ccc}
      \inclps{0.4\textwidth}{!}{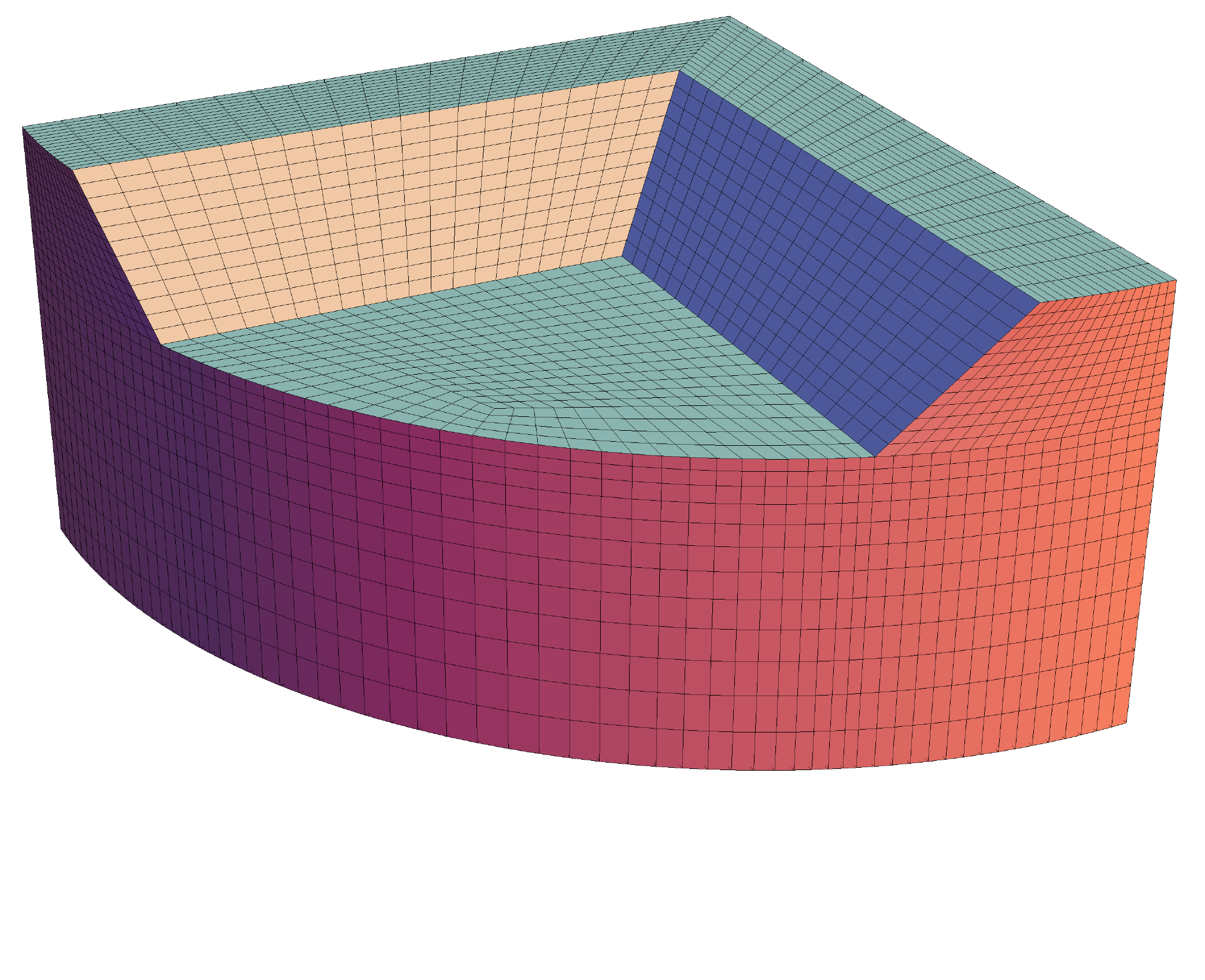} & &
      \inclps{0.4\textwidth}{!}{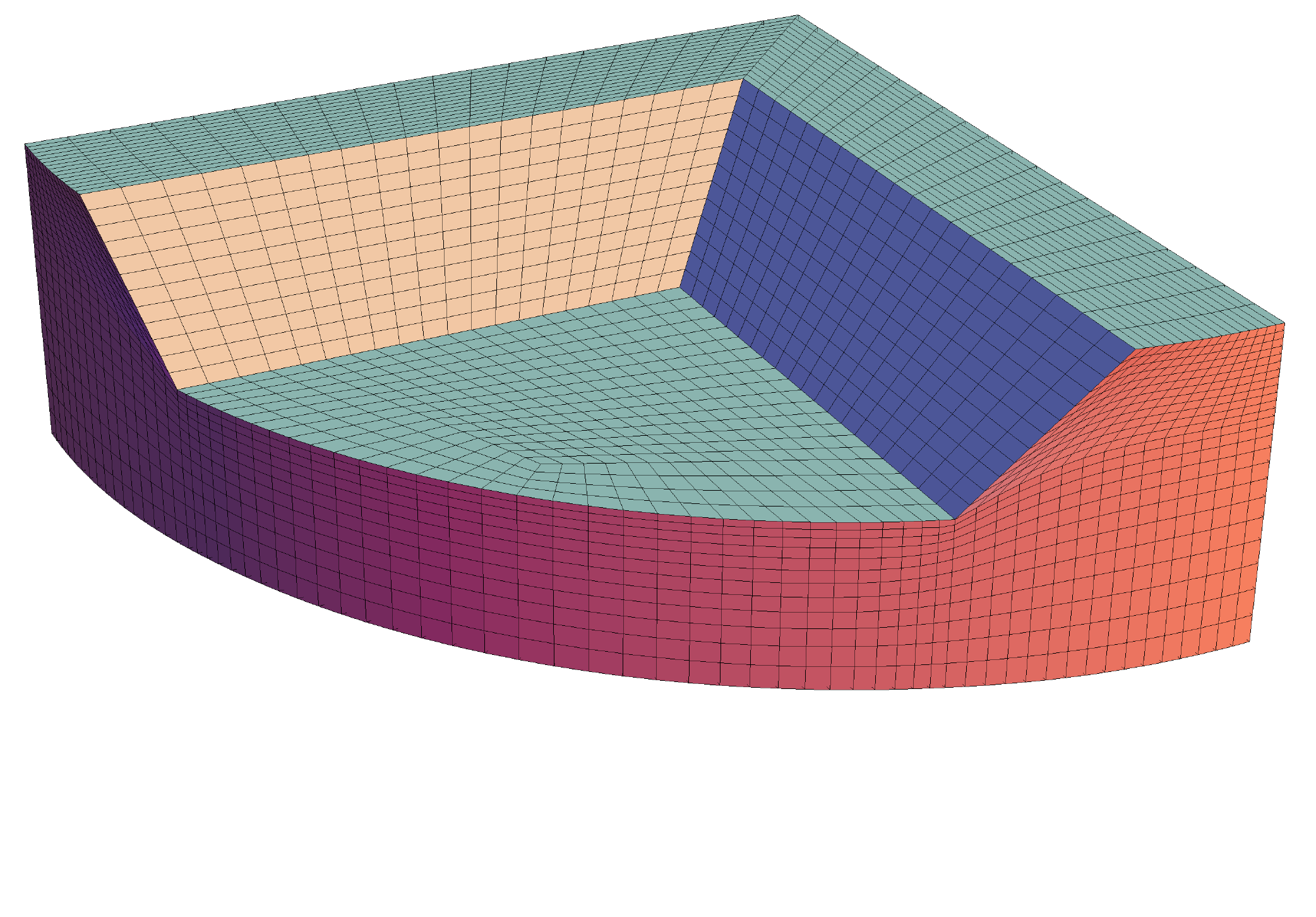}
    \end{tabular}
    }
  \caption{Finite element mesh (one quarter) for the cross-shaped specimen:  undeformed (left) and deformed (right).
           \label{fig:c:1}}
\end{figure}

\begin{figure}[!htcb]
  \centerline{
    \begin{tabular}{cc}
      \inclps{0.5\textwidth}{!}{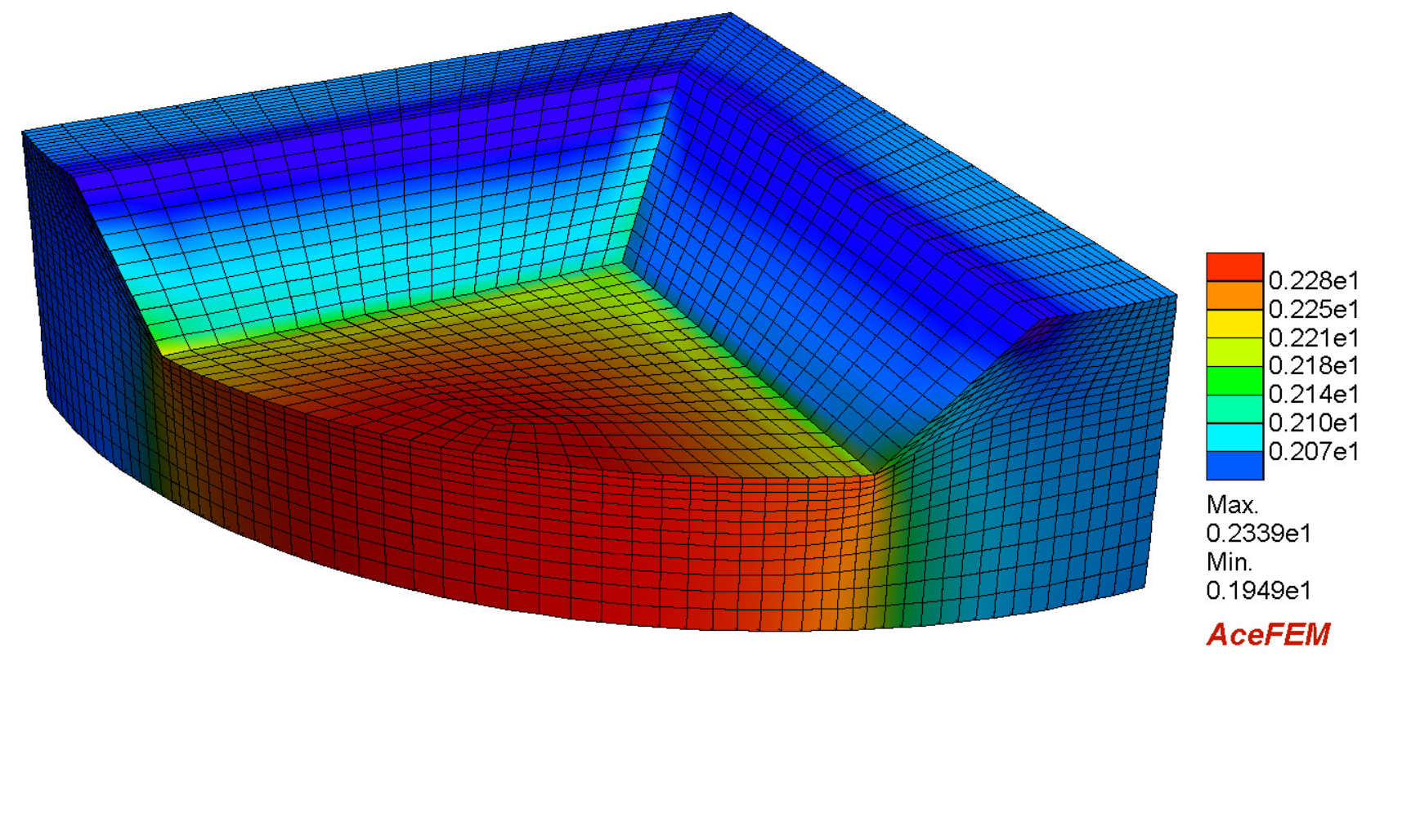} &
      \inclps{0.5\textwidth}{!}{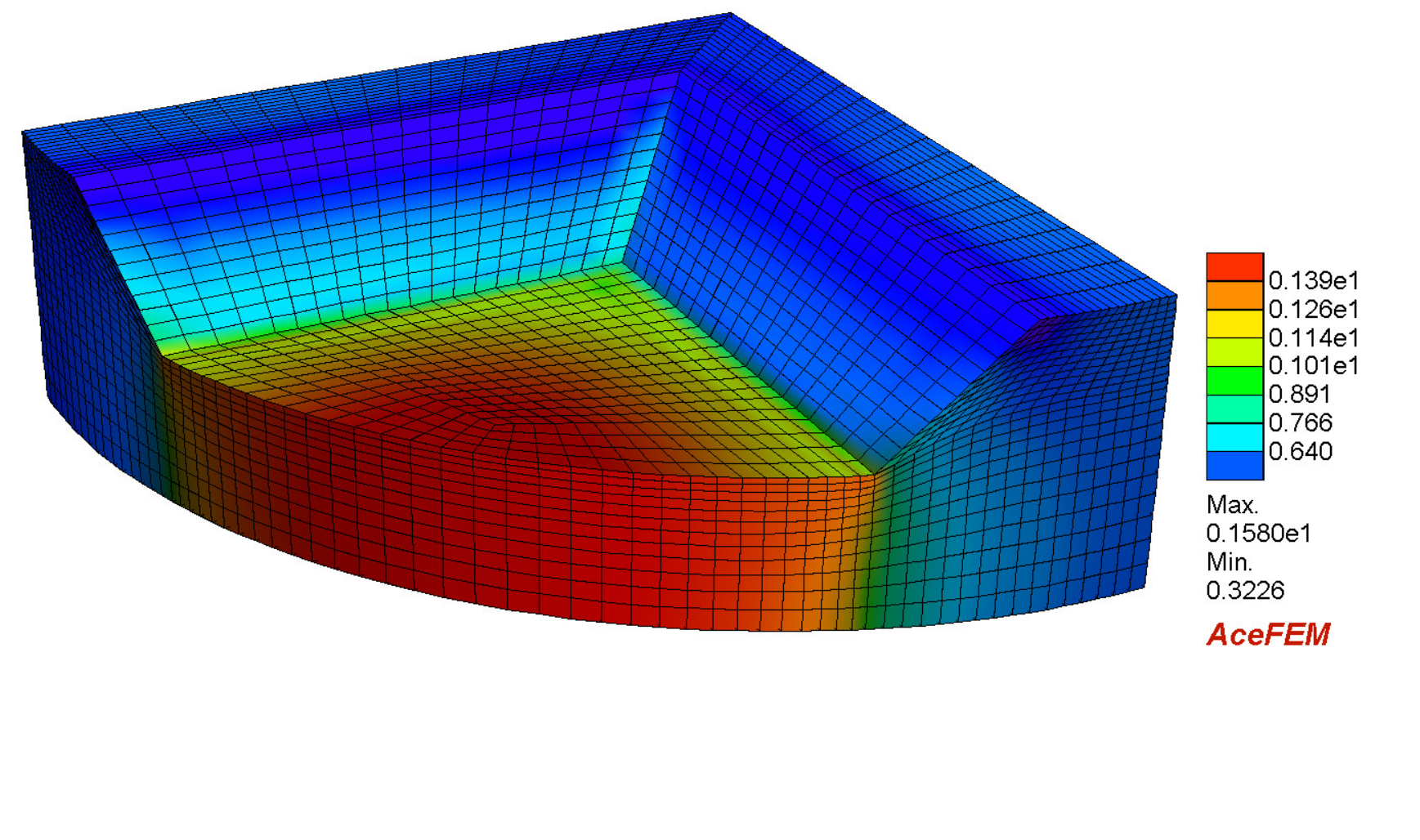}
    \end{tabular}
    }
  \caption{
Simulated distributions of density $\varrho$ (in g/cm$^3$, left) and cohesion $c$ (in MPa, right) within the cross-shaped specimen.
           \label{fig:c:2}}
\end{figure}

\begin{figure}[!htcb]
  \centerline{
    \begin{tabular}{cc}
      \inclps{0.5\textwidth}{!}{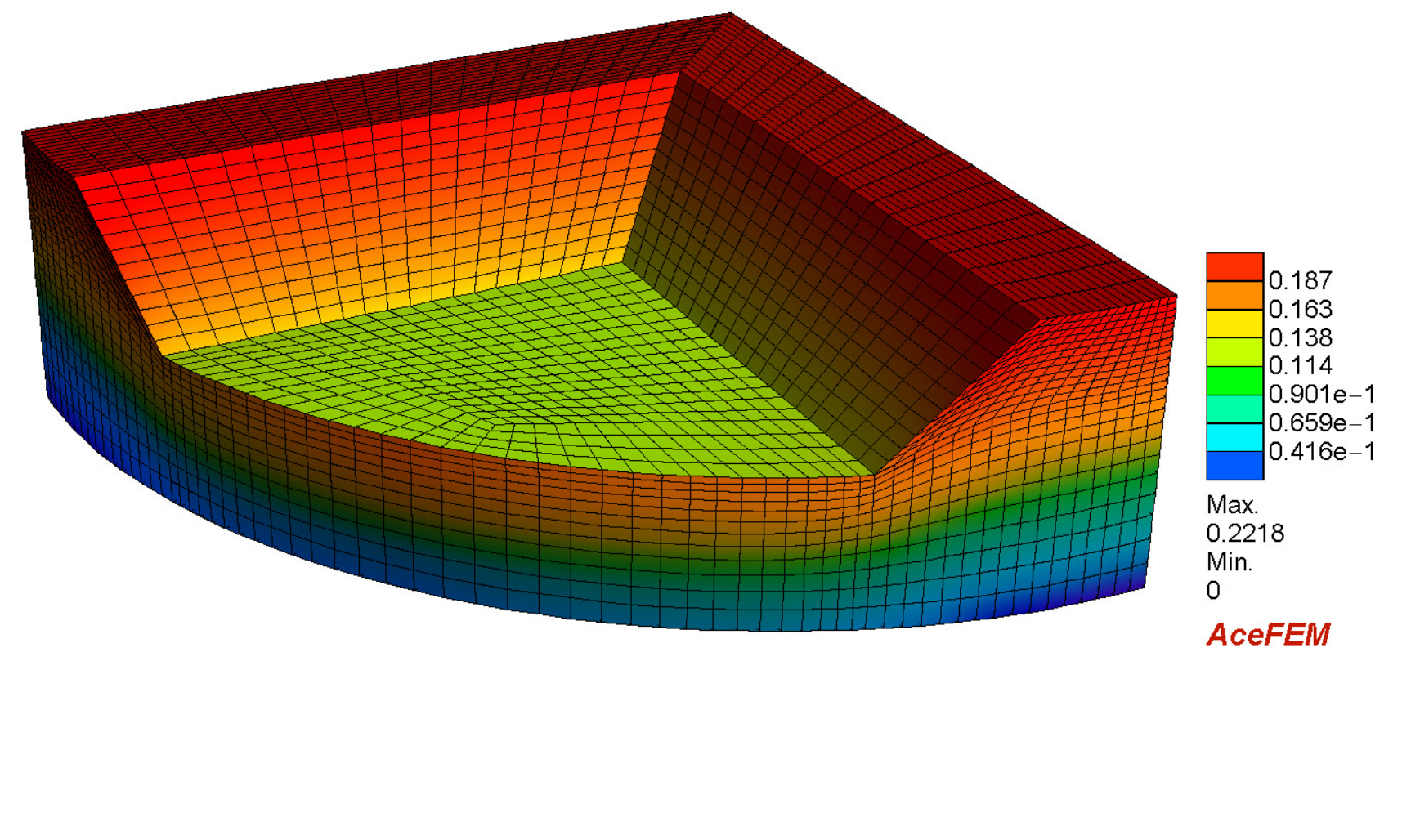} &
      \inclps{0.5\textwidth}{!}{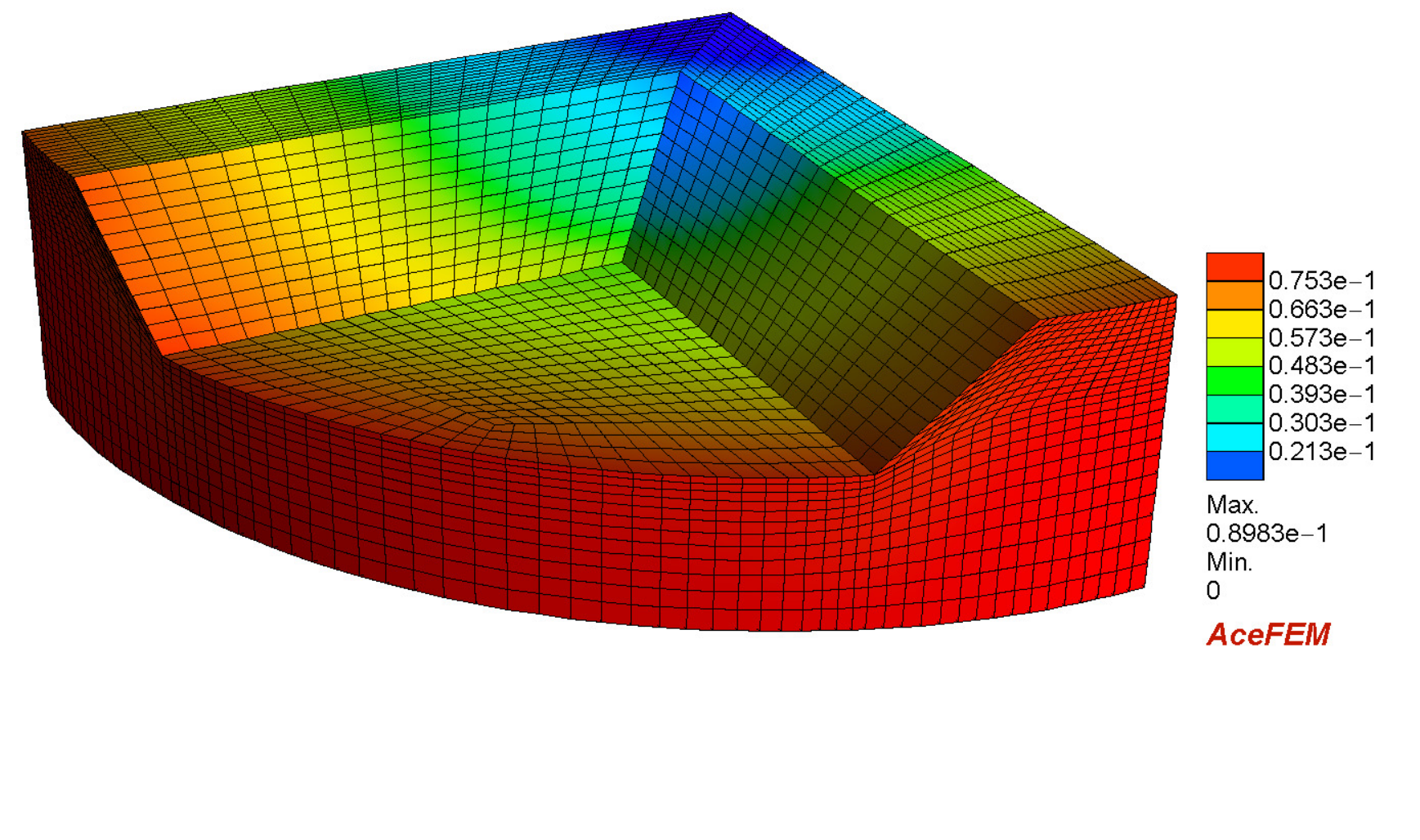}
    \end{tabular}
    }
  \caption{
Simulated spring-back displacement components (in mm) within the cross-shaped specimen:
           axial (left) and radial (right).
           \label{fig:c:3}}
\end{figure}

\begin{figure}[!htcb]
  \centerline{
    \begin{tabular}{cc}
      \inclps{0.5\textwidth}{!}{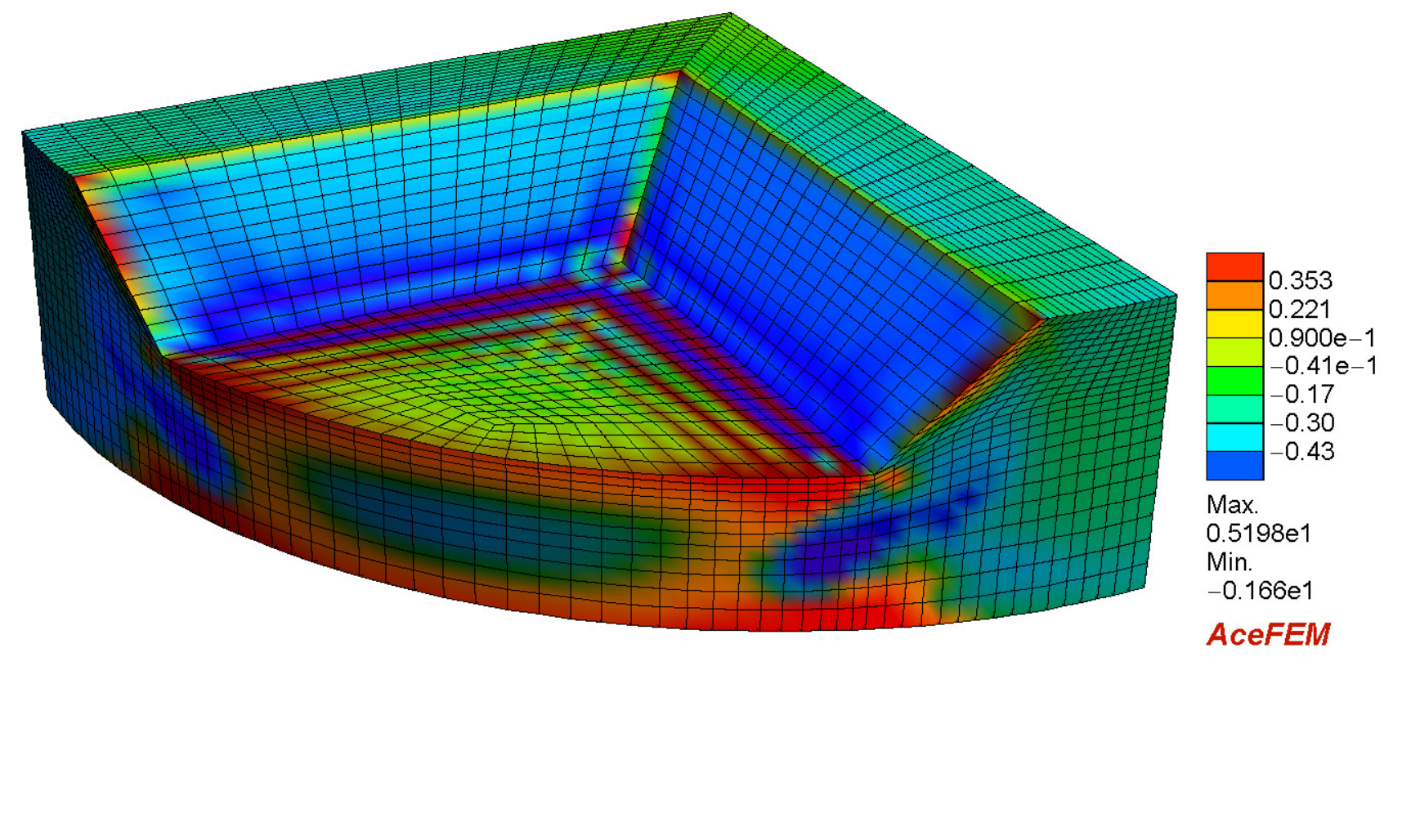} &
      \inclps{0.5\textwidth}{!}{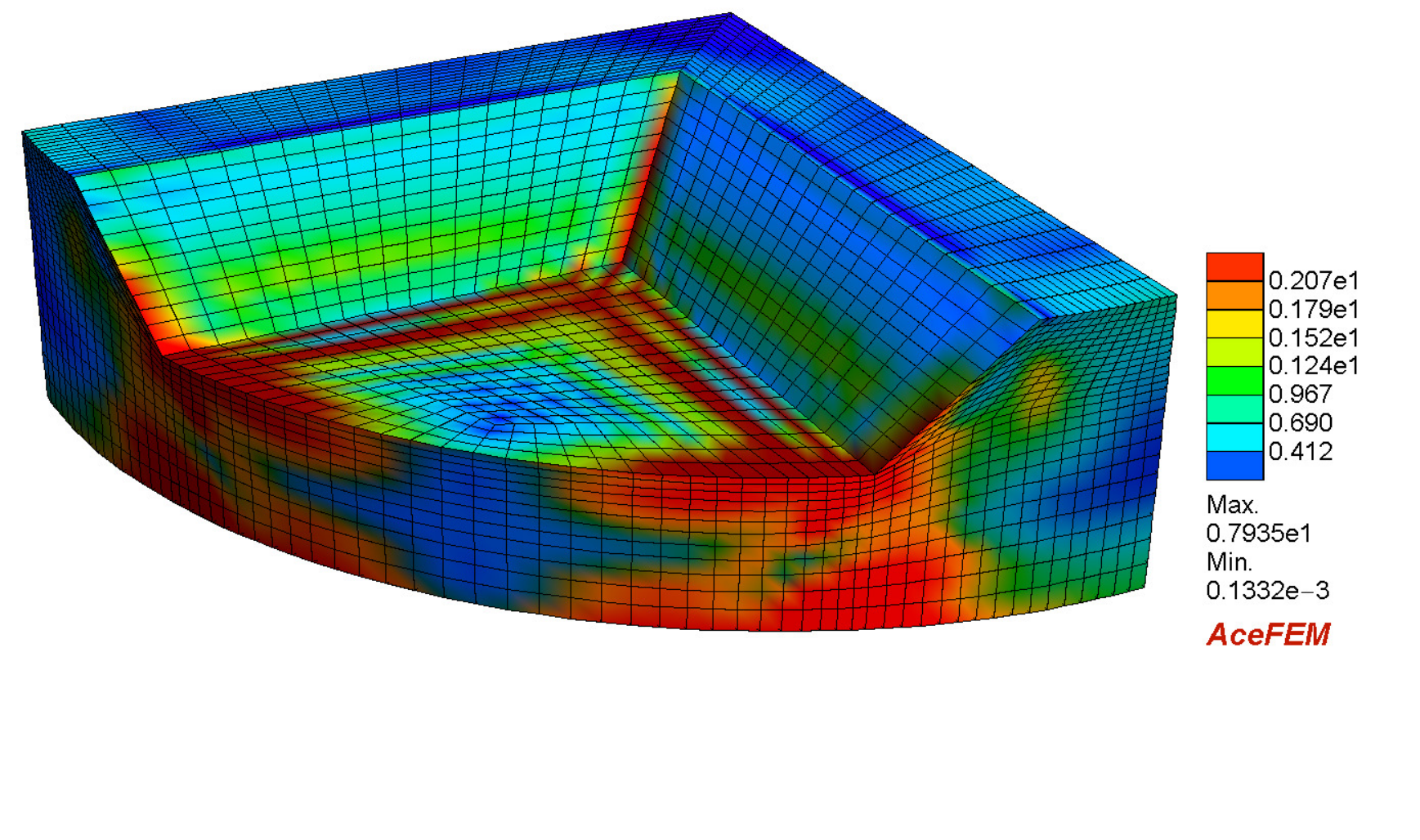}
    \end{tabular}
    }
  \caption{Simulated distributions of residual stresses (in MPa) within the cross-shaped specimen after spring-back:
mean stress $p$ (left) and deviatoric invariant $q$ (right).
           \label{fig:c:4}}
\end{figure}

The undeformed and deformed meshes employed for the simulations are shown in Fig.\
\ref{fig:c:1}, while density and cohesion distribution, spring-back displacements
and residual stresses are shown in Figs.\ \ref{fig:c:2}--\ref{fig:c:4}.
Inhomogeneity of the microstructure of the formed piece is clearly visible
in Fig.\ \ref{fig:c:2} which results in nonuniform deformation during unloading and
spring-back (Fig.\ \ref{fig:c:3}) and significant residual stresses after unloading
(Fig.\ \ref{fig:c:4}).


\begin{figure}[!htcb]
  \centerline{\inclps{0.5\textwidth}{!}{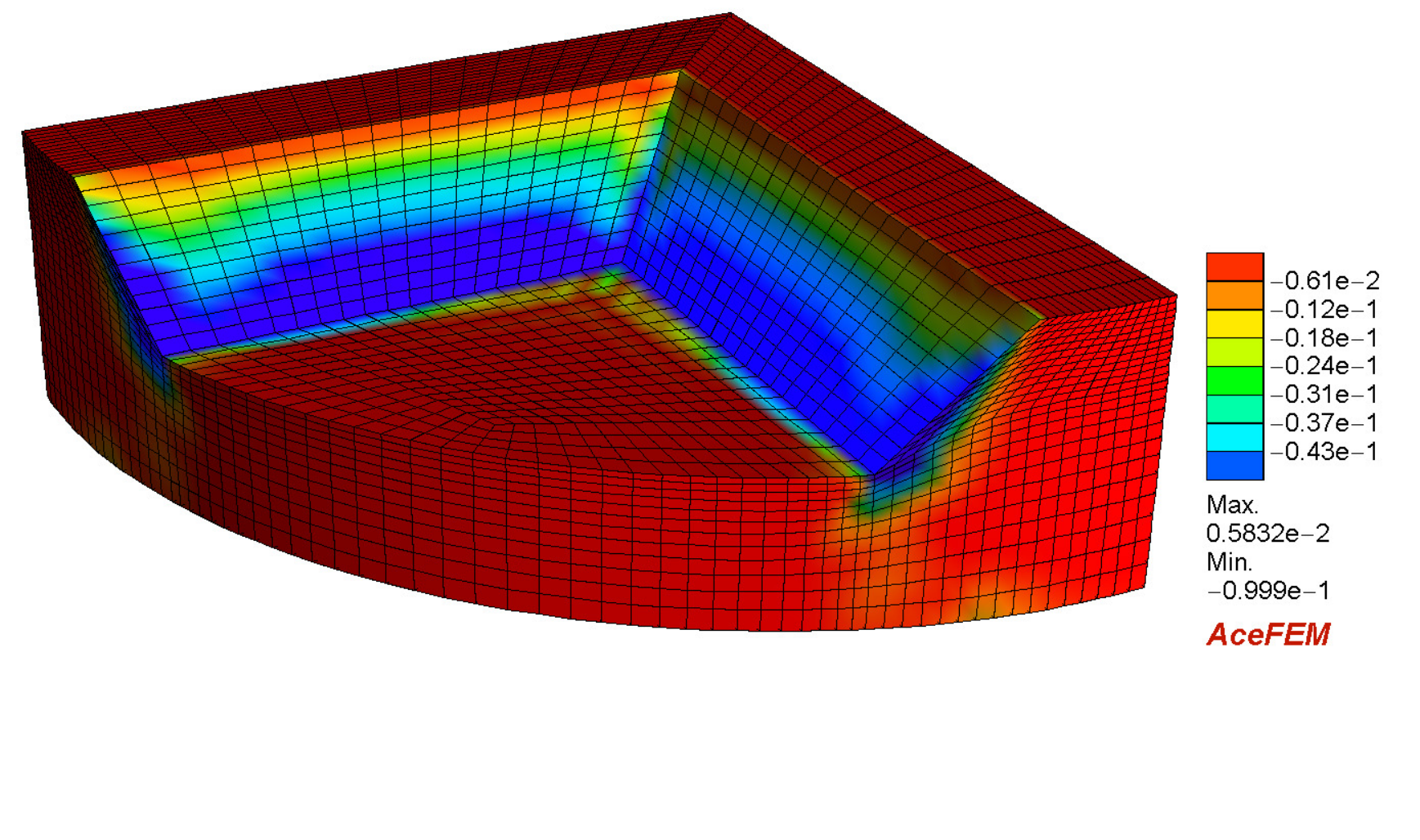}}
  \vspace{-4ex}
  \caption{Simulated map of
change in cohesion $\Delta c$ (in MPa) after spring-back for the cross-shaped specimen. Note the strong decrease near the edge of the grooves.
           \label{fig:c:5}}
\end{figure}

The effect of plastic deformations that accompany spring-back is
illustrated in Fig.\ \ref{fig:c:5}, showing the distribution of the associated
increment of cohesion $\Delta c$. There is a decrease of cohesion (so that the increment is negative) loacalized near the edge of the grooves, which is connected to a plastic softening that may lead to damage of
the piece, as indeed observed in experiment, see upper part of Fig.\ \ref{intatta}.
Softening during spring-back may also be responsible for the
mesh-dependent oscillations of residual stresses, as seen in Fig.\ \ref{fig:c:4}.
Such features of the stress field are not observed during the loading stage.

\begin{figure}[!htcb]
  \centerline{\inclps{0.7\textwidth}{!}{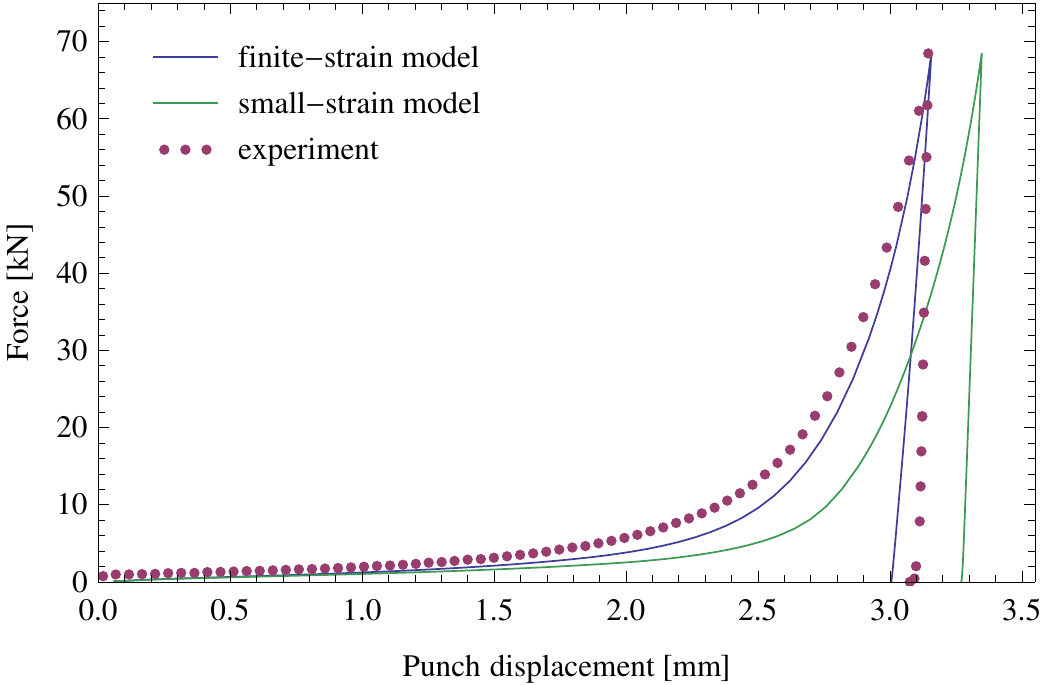}}
  \caption{Simulated (small and large strain analyses) force-displacement curve for the forming of the cross-shaped specimen, compared to experimental results.
           \label{fig:c:6}}
\end{figure}

Finally the simulated force versus displacement relation during forming of the
cross-shaped sample has been compared to the experimental values in Fig.\
\ref{fig:c:6} for the small and large strain cases.

The predictive capabilities of the model, particularly in the large strain version,
are evident from this figure. It is reminded here that the two models
predict a very similar response for uniform compression. The discrepancy between the
two models, which is clearly visible in Fig.\ \ref{fig:c:6}, is thus associated with
non-uniform deformation within the specimen. The configuration changes during powder
forming are large, and, as expected, the finite deformation effects are
significant.


\newpage

\section{Conclusions}

It has been shown that a constitutive model capable of describing compaction of ceramic powders
and based on the concept of elastoplastic coupling can be implemented in an efficient way, both in a small strain and a large strain version, and can be coupled with Coulomb friction at the mould/powder interface. This model allows the simulation of ceramic forming processes and the correct prediction of the spring-back after mould ejection and of the distributions in the green body of: (i.) density, (ii.) cohesion, (iii.) elastic parameters, and (iv.) residual stress.
Therefore, the developed constitutive model can become an effective tool for design of ceramic pieces.

\vspace*{10mm} \noindent {\sl Acknowledgments } Financial support
from the European FP7 - Intercer2 project (PIAP-GA-2011-286110-INTERCER2) is gratefully acknowledged.

\vspace*{5mm} \noindent

 { \singlespace
}


\begin{thebibliography}{}



\bibitem{ah} Ahzi, S., Asaro, R.J., Parks, D.M., 1993. Application of crystal
    plasticity theory for mechanically processed BSCCO superconductors. Mech. Mater.
    15, 201–222.

\bibitem{ala} Alart P., Curnier A. 1991. A mixed formulation for frictional contact
    problems prone to Newton like solution methods. Comp. Meth. Appl. Mech. Eng. 92,
    353-375.

\bibitem{ar} Ariffin, A.K., Gethin, D.T., Lewis, R.W., 1998. Finite element
    simulation and experimental validation for multilevel powder compact. Powder
    Metall. 41, 189–197.

\bibitem{ay1} Aydin, I., Briscoe, B.J., Sanliturk, K.Y., 1997a. Dimensional variation of die-pressed ceramic green compacts: comparison of a finite element modelling with experiment. J. Eur. Ceram. Soc. 17, 1201–1212.

\bibitem{ay2} Aydin, I., Briscoe, B.J., Ozkan, N., 1997b. Modeling of powder compaction: a review. MRS Bull. 22, 45–51.

\bibitem{bal} Balakrishnan, A., Martin, C.L., Saha, B.P. and Joshi, S. 2011. Modelling of compaction and green strength of aggregated ceramic powders. J. Am. Ceram. Soc. 94, 1046-1052.

\bibitem{ba1} Baklouti, S., Chartier, T., Gault, C., Baumard, J.F. 1997. The Eff?ect of Binders on the Strength and Young's Modulus of Dry Pressed Alumina. J. Europ. Ceramic Soc. 18, 323-328.

\bibitem{ba2} Baklouti, S., Chartier, T., Gault, C., Baumard, J.F. 1999. Young's Modulus of Dry-pressed Ceramics: The Eff?ect of the Binder. J. Europ. Ceramic Soc. 19, 1569-1574.

\bibitem{bigoni} Bigoni, D. 2012. {\it Nonlinear Solid Mechanics. Bifurcation theory and material instability}, Cambridge University Press.

\bibitem{bigoni2} Bigoni, D. and Piccolroaz, A. 2004. Yield criteria for quasibrittle and frictional materials. Int. J. Solids Struct., 41, 2855-2878.

\bibitem{bosi} Bosi, F., A. Piccolroaz, M. Gei, F. Dal Corso, A. Cocquio and Bigoni,
    D. 2013. Experimental investigation of the elastoplastic response of aluminum
    silicate spray-dried powder during cold compaction. J. Europ. Ceramic Soc.
    Submitted.

\bibitem{br1} Brandt, J., Nilsson, L., 1998. FE-simulation of compaction and solid-state sintering of cemented carbides. Mech. Cohesive-Frict. Mater. 3, 181–205.

\bibitem{br2} Brandt, J., Nilsson, L., 1999. A constitutive model for compaction of granular media, with account for deformation induced anisotropy. Mech. Cohesive-Frict. Mater. 4, 391–418.

\bibitem{bran} Brannon, R.M. and Leelavanichkul, S. (2010) Received: A multi-stage return algorithm for solving the classical damage component of constitutive models for rocks, ceramics, and other rock-like media. \IJF 163, 133–149.

\bibitem{ca} Carneim, T.J., Green, D.J.  2001. Mechanical Properties of Dry-Pressed Alumina Green Bodies. J. Am. Ceram. Soc., 84, 1405–1410.

\bibitem{cooper} Cooper, A.R., Eaton, L.E., 1962. Compaction behavior of several ceramic powders. J. Am. Ceram. Soc. 45, 97–101.

\bibitem{dou} Dougill, J.W., 1976. On stable progressively fracturing solids. Z. Angew. Math. Phys. 27, 423–437.

\bibitem{ew1} Ewsuk, K.G., Arg\"uello, J.G., Zeuch, D.H., Farber, B., Carinci, L., Kaniuk, J., Keller, J., Cloutier, C., Gold, B., Cass, R.B., French, J.D., Dinger, B., Blumenthal, W., 2001a. CRADA develops model for powder pressing and die design, Part one. Am. Ceram. Soc. Bull. 80, 53–60.

\bibitem{ew2} Ewsuk, K.G., Arg\"uello, J.G., Zeuch, D.H., Farber, B., Carinci, L., Kaniuk, J., Keller, J., Cloutier, C., Gold, B., Cass, R.B., French, J.D., Dinger, B., Blumenthal, W., 2001b. CRADA develops model for powder pressing and die design, Part two. Am. Ceram. Soc. Bull. 80, 41–46.

\bibitem{gu} Gu, Y., Henderson, R.J., and Chandler, H.W. 2006. Visualizing isostatic pressing of ceramic powders using finite element analysis. J. Europ. Ceramic Soc. 26, 2265-2272.

\bibitem{hen} Henderson, R.J., Chandler, H.W., Akisanya, A.R., Barber, H. and Moriarty, B. 2000. Finite element modelling of cold isostatic pressing. J. Europ. Ceramic Soc. 20, 1121--1128.

\bibitem{hue1}  Hueckel, T., 1975. On plastic flow of granular and rock-like materials with variable elasticity moduli. Bull. Pol. Acad. Sci., Ser. Techn. 23, 405–414.

\bibitem{hue2} Hueckel, T., 1976. Coupling of elastic and plastic deformation of bulk solids. Meccanica 11, 227–235.

\bibitem{keller} Keller, J.M., French, J.D., Dinger, B., McDonough, M., Gold, B., Cloutier, C., Carinci, L., Van Horn, E., Ewsuk, K., Blumenthal, B., 1998. Industry, government team to improve ceramic manufacturing. Am. Ceram. Soc. Bull. 77, 52–57.

\bibitem{ki1}  Kim, H.G., Gillia, O., Dor\'emus, P., Bouvard, D., 2002. Near net shape processing of a sintered alumina component: adjustment of pressing
parameters through finite element simulation. Int. J. Mech. Sci. 44, 2523–2539.

\bibitem{ki} Kim, H.G., Lee, H.M., Kim, K.T. 2001. Near-net-shape forming of ceramic
    powder under cold combination pressing and pressureless sintering. J. Eng. Mat.
    Tech. 123, 221-228.


\bibitem{ko1} Korelc, J., 2002. Multi-language and multi-environment generation of nonlinear finite element codes. Engineering with Computers, 18, 312–327.


\bibitem{ko2} Korelc, J., 2009. Automation of primal and sensitivity analysis of transient coupled problems. Comp. Mech., 44, 631–649.

\bibitem{ko} Kounga Njiwa, A.B., Aulbach, E., R\"odel, J.  2006. Mechanical Properties of Dry-Pressed Powder Compacts: Case Study on Alumina Nanoparticles. J. Am. Ceram. Soc., 89, 2641–2644.

\bibitem{lee} Lee, S.C., Kim, K.T. 2008. Densification behaviour of nanocrystalline titania powder under cold compaction. Powder Tech. 186, 99-106.

\bibitem{len} Lengiewicz, J., Korelc, J., Stupkiewicz, S. 2011. Automation of
    finite element formulations for large deformation contact problems. Int. J. Num.
    Meth. Eng. 85, 1252-1279.

\bibitem{par} Park, H., Kim, K.T. 2001. Consolidation behaviour of SiC powder under
    cold compaction. Mat. Sci. Eng. A 299, 116-124.


\bibitem{penasa} Penasa, M., Piccolroaz, A., Argani, L. and Bigoni, D. 2013.
Integration algorithms of elastoplasticity for ceramic powder compaction. J. Europ. Ceramic Soc. Submitted.


\bibitem{pic1} Piccolroaz, A., Bigoni, D., 2009. Yield criteria for quasibrittle and frictional materials: a generalization to surfaces with corners.
Int. J. Solids Struct., 46, 3587-3596.

\bibitem{pic2} Piccolroaz, A., Bigoni, D., Gajo, A. 2006a. An elastoplastic framework for granular materials becoming cohesive through mechanical densification. Part I - small strain formulation. Europ. J. Mech. A: Solids, 25, 334-357.

\bibitem{pic3} Piccolroaz, A., Bigoni, D., Gajo, A. 2006b. An elastoplastic framework for granular materials becoming cohesive through mechanical densification. Part II - the formulation of elastoplastic coupling at large strain.
Europ. J. Mech. A: Solids, 25, 358-369.

\bibitem{pic7} Piccolroaz, A., Penasa, M., Argani, L., Bigoni, D. 2013. Integration algorithms of elastoplasticity for ceramic powder compaction. Submitted.

\bibitem{stup} Stupkiewicz, S., Denzer, R.P., Piccolroaz, A., Bigoni, D. 2013.
Implicit yield function formulation for granular and rock-like materials. Submitted.

\bibitem{ze}  Zeuch, D.H., Grazier, J.M., Arg\"uello, J.G., Ewsuk, K.G., 2001. Mechanical properties and shear failure surfaces for two alumina powders in triaxial compression. J. Mater. Sci. 36, 2911–2924.

\bibitem{zip} Zipse, H., 1997. Finite-element simulation of die pressing and sintering of a ceramic component. J. Eur. Ceram. Soc. 17, 1707–1713.



\end{thebibliography}
\end{document}